\PassOptionsToPackage{unicode}{hyperref}
\PassOptionsToPackage{hyphens}{url}
\documentclass[%
reprint,
amsmath,amssymb,
aps,
]{revtex4-2}
\usepackage{filecontents}
\usepackage{etoolbox}
\usepackage{graphicx}
\usepackage{dcolumn}
\usepackage{bm}
\usepackage{svg}
\usepackage{booktabs}
\usepackage{multirow}
\usepackage{titlesec}
\titleformat{\section}[hang]{\normalfont\bfseries}{\thesection}{12em}{}
\usepackage{colortbl}

\usepackage{pdfpages}
\usepackage{pgffor} 

\makeatletter
\AtBeginDocument{\let\LS@rot\@undefined}
\patchcmd{\@makecaption}{\@ifdim{\wd\@tempboxa >\hsize}}{\@firstoftwo}{}{}
\patchcmd{\@outputpage@head}{\@ifx{\LS@rot\@undefined}{}{\LS@rot}}{}{}{}
\makeatother

\def\supplementfilename{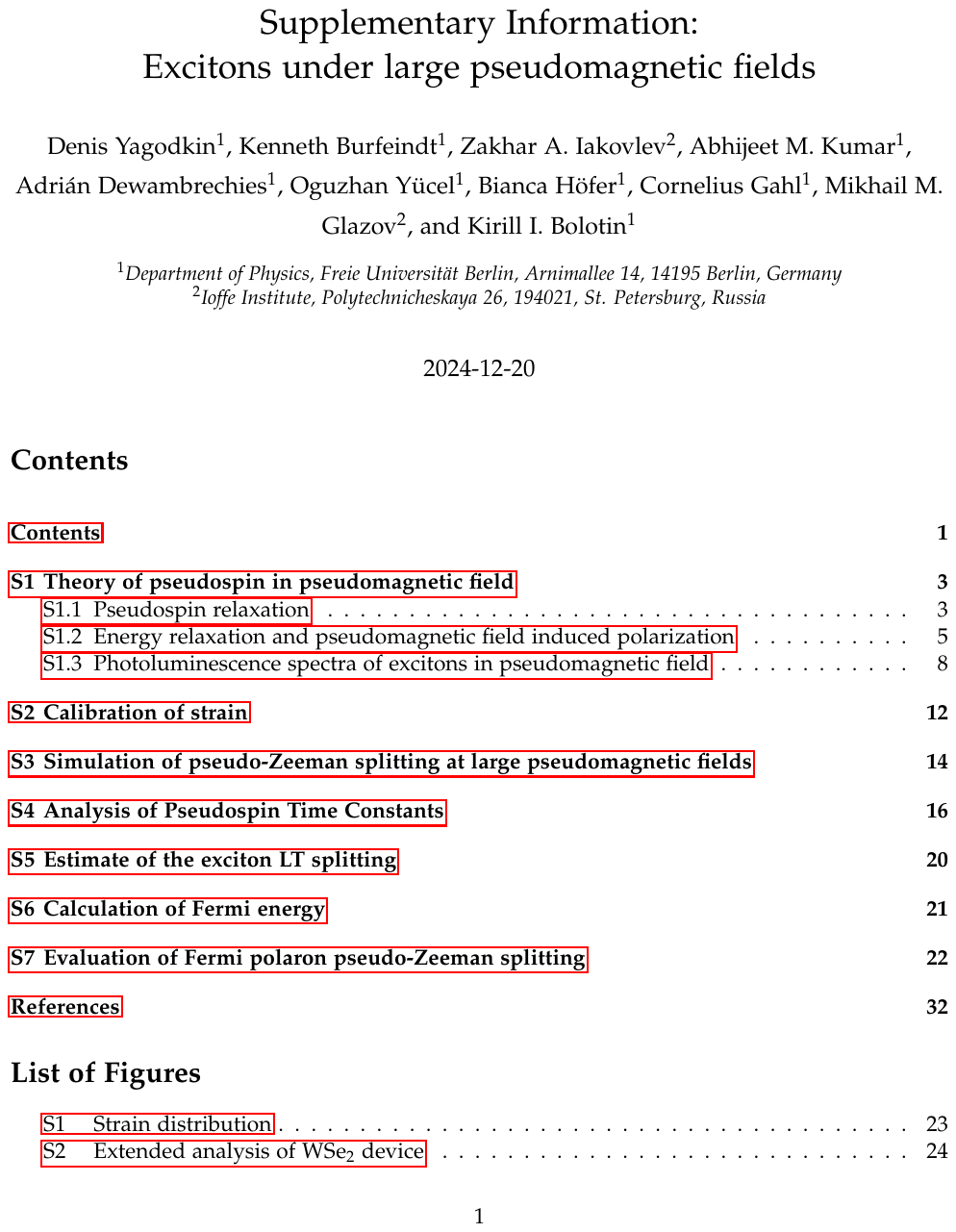}

\pdfximage{\supplementfilename}
\def\numbersupplementpages{\the\pdflastximagepages}

\newif\ifarXiv
\arXivtrue 

\usepackage{color}
\usepackage{xcolor}
\usepackage[normalem]{ulem}

\begin{document}

\title{Excitons under large pseudomagnetic fields}

\author{Denis Yagodkin$^1$}
\author{Kenneth Burfeindt$^1$}
\author{Zakhar A. Iakovlev$^2$}
\author{Abhijeet M. Kumar$^1$}
\author{Adrián Dewambrechies$^1$}
\author{Oguzhan Yücel$^1$}
\author{Bianca Höfer$^1$}
\author{Cornelius Gahl$^1$}
\author{Mikhail M. Glazov$^2$}
\author{Kirill I. Bolotin$^1$}
\affiliation{$^1$Department of Physics, Freie Universitat Berlin, Arnimallee 14, 14195 Berlin, Germany}
\affiliation{$^2$Ioffe Institute, Polytechnicheskaya 26, 194021, St. Petersburg, Russia}
\date{\today}

\begin{abstract}
  \textbf{Abstract}: Excitons in Transition Metal Dichalcogenides (TMDs) acquire a spin-like quantum number, a pseudospin, originating from the crystal’s discrete rotational symmetry. Here, we break this symmetry using a tunable uniaxial strain, effectively generating a pseudomagnetic field exceeding 40~Tesla. Under this large field, we demonstrate pseudospin analogs of spintronic phenomena such as the Zeeman effect and Larmor precession. Moreover, we determine previously inaccessible fundamental properties of TMDs, including the strength of the depolarizing field responsible for the loss of exciton coherence. Finally, we uncover the bosonic -- as opposed to fermionic --  nature of many-body excitonic species using the pseudomagnetic equivalent of the $g$-factor spectroscopy. Our work is the first step toward establishing this spectroscopy as a universal method for probing correlated many-body states and realizing pseudospin analogs of spintronic devices. 
\end{abstract} 

\maketitle
\begin{center}
\textbf{Introduction}
\end{center}
The coupling between an electron's spin and magnetic field is the source of omnipresent phenomena ranging from the Zeeman effect, the Larmor effect, and magnetic resonances to anomalous and quantum spin Hall effects. The counterparts of these phenomena also arise in non-magnetic systems with two degenerate but distinct states, requiring a new quantum number known as pseudospin to distinguish them~\cite{glazov_exciton_2022, yu_dirac_2014,glazovSpinTransportEffects2010}. The dynamics of this pseudospin mirror those of a spin in a magnetic field when degeneracy is lifted by an external perturbation that acts as a ``pseudomagnetic field"~\cite{ilanPseudoelectromagneticFields3D2020}. For example, the polarization of light in photonic crystals can be treated as a pseudospin, with optical birefringence playing the role of a pseudomagnetic field~\cite{rechcinskaEngineeringSpinorbitSynthetic2019,renNontrivialBandGeometry2021a}. The layer of charge carrier localization in bilayer graphene can also be considered as a pseudospin, with the out-of-plane electric field playing the role of a pseudomagnetic field~\cite{zhangSpontaneousQuantumHall2011}. These conceptual parallels enabled realization of flat bands in photonic crystals~\cite{barsukova_direct_2024,barczykObservationLandauLevels2024} and unconventional superconductivity in twisted bilayer of graphene~\cite{khalafChargedSkyrmionsTopological2021,bistritzerMoireBandsTwisted2011, ohEvidenceUnconventionalSuperconductivity2021, haoElectricFieldTunable2021}.

\begin{figure*}
  \includegraphics[width=0.95\linewidth]{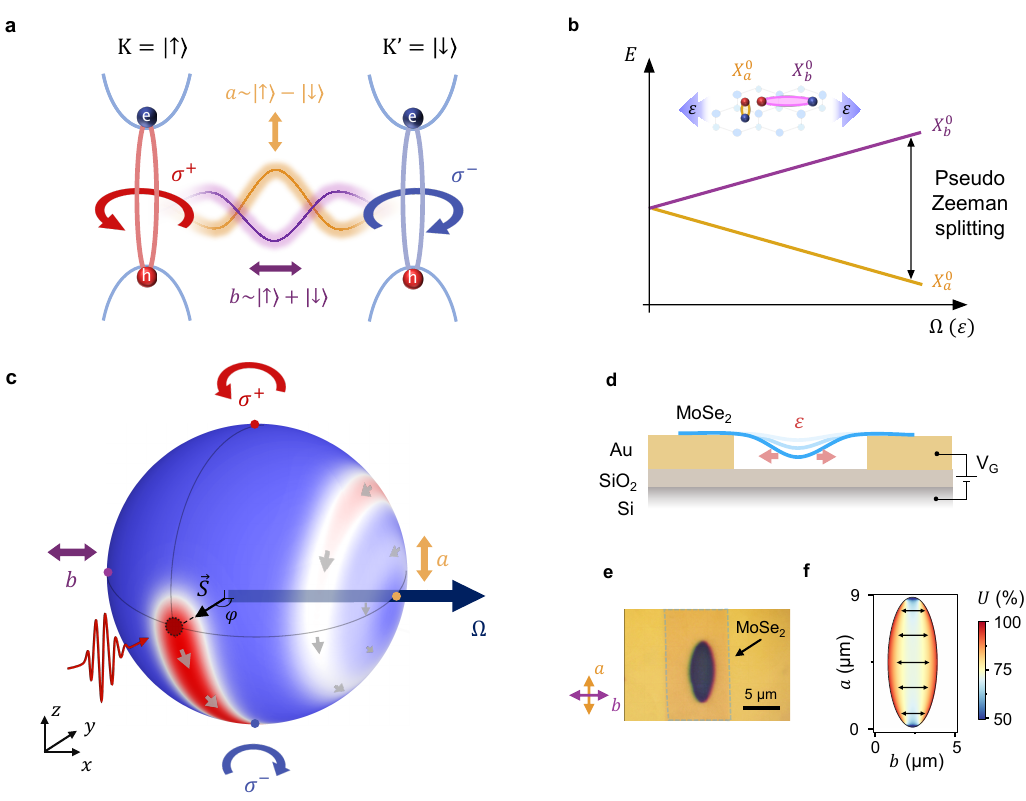}    
  \caption{
    \textbf{Excitons, strain, and pseudomagnetic field. a)} Different superpositions of excitons in K and K' valleys are excited by light with distinct polarizations. Circularly polarized light, $\sigma^+$ or $\sigma^-$, couples to K or K' excitons, respectively (red and blue arrows), whereas linearly polarized light (purple and orange arrows) generates superpositions of these excitons.
    \textbf{b)} The energy degeneracy between the excitons generated by light with two orthogonal linear polarizations is lifted in the presence of uniaxial strain \(\varepsilon\).
    \textbf{c)} Bloch sphere representation of pseudospin. Each coherent superposition of K and K' excitons corresponds to a pseudospin vector $\bm S$ on the Bloch sphere. The $\sigma^+$ or $\sigma^-$ circularly polarized light couples to the states at the poles, while linearly polarized light excites the states in the equatorial plane. In the presence of uniaxial strain, \(\bm{S}\) undergoes damped Larmor-like precession around the strain-induced pseudomagnetic field \(\bm{\Omega}\).
    \textbf{d)} Straining technique: an applied gate voltage (\(V_{G}\)) induces tensile strain \(\varepsilon\) (pink arrows) in suspended MoSe$_2$ or WSe$_2$ monolayer (blue) via electrostatic force.
    \textbf{e)} Optical image of a suspended MoSe$_2$ monolayer.
    \textbf{f)} COMSOL simulation of strain uniaxiality \(U\) in a typical device.
  }
  \label{fig:1}
\end{figure*}

One especially versatile system for exploring pseudomagnetic phenomena is monolayers of Transition Metal Dichalcogenides (TMDs)~\cite{wangColloquiumExcitonsAtomically2018,manzeli_2d_2017,xu_spin_2014}. There, a broken inversion symmetry gives rise to degenerate valleys at K and K' points at the corners of the Brillouin zone, that host tightly bound excitons (Fig.~\ref{fig:1}a). The pseudospin associated with this degeneracy can be initialized and read out optically: $\sigma^+$ ($\sigma^-$) polarized light couples to excitons at the K (K') valleys (pseudospin up and down, respectively), while linear polarization couples to a superposition of K and K' excitons, corresponding to an in-plane pseudospin~\cite{yu_dirac_2014, glazov_exciton_2014,wangColloquiumExcitonsAtomically2018,manzeli_2d_2017}.

Recently, it has been suggested that homogeneous uniaxial mechanical strain can be used to generate pseudomagnetic fields in TMDs~\cite{glazov_exciton_2022, yu_dirac_2014,xu_spin_2014}. Such a strain lifts the energy degeneracy between the excitons coupled to light polarized along (X\(^0_{b}\)) and perpendicular (X\(^0_{a}\)) to the strain axis (Fig.~\ref{fig:1}b). Therefore, its influence can be described as an in-plane pseudomagnetic field. Critically, the effects arising from this field should extend beyond conventional magnetic phenomena. First, the strength of a pseudomagnetic field can exceed 300~T~\cite{bertolazzi_stretching_2011, carrascosoStrainEngineeringSingle2021a, glazov_coherent_2022,lloyd_band_2016}, leading to the emergence of new physical regimes~\cite{maoEvidenceFlatBands2020,shiLargeareaPeriodicTunable2020, niggeRoomTemperatureStraininduced2019, barsukova_direct_2024}. Second, various states in TMDs~---~including charged, dark, and intervalley excitons~---~feature complex pseudospin compositions, leading to intricate many-body interactions~\cite{heValleyPhononsExciton2020}. Finally, the coupling between pseudospin and momentum degrees of freedom differs from the familiar spin–orbit coupling, potentially resulting in new types of Hall effects~\cite{guineaEnergyGapsZerofield2010, pengStrainEngineering2D2020a}.

Previously, strain-related pseudomagnetic fields acting on the electrons valley pseudospin have been demonstrated in monolayer graphene~\cite{levyStrainInducedPseudoMagnetic2010,georgi2017tuning}, but such a field in graphene requires a highly inhomogeneous strain profile and arises only at a nanometer scale. Additionally, optical Stark effect has been used to generate pseudomagnetic fields in TMDs~\cite{ye_optical_2017, kimUltrafastGenerationPseudomagnetic2014a}; however, this approach is not suitable for studying time-independent optical and transport phenomena, resolving closely lying states, and probing low oscillator strength excitons. Despite the strong potential of strain-induced pseudomagnetic fields in TMDs, their experimental exploration remains challenging. This is due to the complexity of mechanically manipulating TMDs at cryogenic temperatures, which is critical to avoid scattering-induced loss of pseudospin~\cite{glazov_exciton_2014, glazov_coherent_2022}. Furthermore, a large and controlled uniaxial strain ($> 0.5\%$) is required to resolve the pseudomagnetic shift of excitonic lines comparable to the excitonic linewidth~\cite{glazov_exciton_2022, mitioglu_anomalous_2019, mitioglu_observation_2018}.

These challenges leave several questions unanswered. First, what analogs of conventional magnetic phenomena carry over to exciton valley  pseudospins? Second, what fundamental properties of TMDs define pseudospin dynamics? Finally, how do many-body excitonic states with complex valley character (e.g., charged excitons, biexcitons) respond to pseudomagnetic fields, and could the ``$g$-factors" of these states provide insights into their nature? To tackle these questions, we developed a platform to control exciton valley pseudospin at cryogenic temperatures using a strain-induced pseudomagnetic field in TMDs. We study the emergent analogs of magnetic phenomena and establish pseudomagnetic $g$-factor fingerprinting as a technique to investigate the nature of many-body states.
\begin{center}
  \textbf{Results}
\end{center}
\noindent \textbf{Pseudospin in a strained TMD.} The spatial symmetry of TMDs dictates that a linearly polarized photon in a state \(\alpha|\sigma^{+}\rangle+\beta|\sigma^{-}\rangle\), with \(|\alpha|^{2} = |\beta|^{2} = 1/2\), creates a coherent superposition of bright excitons with wavefunctions residing in K and K' valleys, \(\Psi = \alpha\left| X_{\text{KK}} \right\rangle + \beta\left| X_{\text{K'K'}} \right\rangle\). The spinor \(\chi = (\alpha,\ \beta)\) then determines the pseudospin \(\bm{S}\) in the same way as the electron spin is defined in quantum mechanics: \(\bm{S} = \left( \operatorname{Re}(\alpha\beta^{*}),\ \operatorname{Im}(\alpha^{*}\beta),\ |\alpha|^{2} - |\beta|^{2} \right)\). The application of mechanical strain breaks the underlying symmetries of TMDs, thereby affecting the pseudospin degree of freedom~\cite{glazov_exciton_2022, yu_dirac_2014}. The effect of strain on the exciton's pseudospin in the limit of zero exciton momentum is described by the following Hamiltonian:
\begin{equation}\label{eq:1}
  \scalebox{0.97}{$H = \begin{bmatrix} \frac{A}{2}\left(\varepsilon_{xx} + \varepsilon_{yy}\right) & \frac{B}{2}\left(\varepsilon_{xx} - \varepsilon_{yy} -2\mathrm  i \varepsilon_{xy}\right) \\ \frac{B}{2}\left(\varepsilon_{xx} - \varepsilon_{yy} +2\mathrm  i \varepsilon_{xy}\right)  & \frac{A}{2}\left(\varepsilon_{xx} + \varepsilon_{yy}\right) \end{bmatrix}$},
\end{equation}
where \(\varepsilon_{xx},\ \varepsilon_{yy},\ \varepsilon_{xy} = \varepsilon_{yx}\) are the components of the strain tensor, and \(A, B\) are material-specific parameters. The diagonal terms describe the well-known energy shift of the excitons under biaxial strain at a rate \(A \approx -100\)~meV/\%~\cite{hernandezlopezStrainControlHybridization2022, kumar_strain_2024, carrascosoStrainEngineeringSingle2021a}. It is evident that KK and K'K' excitons, related by time-reversal symmetry, always remain energetically degenerate. However, the off-diagonal terms suggest that an application of uniaxial (\(\varepsilon_{xx} \neq \varepsilon_{yy}\)) or shear (\(\varepsilon_{xy} \neq 0\)) strain \emph{mixes} excitons in K and K' valleys. This effect becomes apparent if we rearrange the Hamiltonian in the form
\(
  H =  H_{0} + \frac{\hbar}{2}\left( \bm{\Omega} \cdot \bm{\sigma} \right),
\)
where \(H_0=A\left(\varepsilon_{xx} + \varepsilon_{yy} \right)\sigma_0/2\) is the diagonal part of Eq.~(\eqref{eq:1}),  \(\bm{\Omega} = (B/\hbar)(\varepsilon_{xx} - \varepsilon_{yy},\ 2\varepsilon_{xy},\ 0)\), \(\sigma_0\) is the identity matrix, and \(\bm{\sigma} = (\sigma_{x},\ \sigma_{y},\ \sigma_{z})\) is the vector of Pauli matrices acting in the pseudospin basis. This Hamiltonian is formally equivalent to that of a spin in a magnetic field, with the vector $\bm \Omega$ playing the role of the pseudomagnetic field (Note~S1). We therefore expect the presence of analogs of magnetic phenomena in strained devices.

\noindent\textbf{Generation of pseudomagnetic field and detection of pseudospin.} To experimentally probe the strain-induced pseudomagnetic phenomena, we combine nanomechanics with optical spectroscopy. We generate a strong pseudomagnetic field at cryogenic temperatures using a technique based on tensioning of a suspended monolayer with electrostatic force (Fig.~\ref{fig:1}d) that we recently developed~\cite{hernandezlopezStrainControlHybridization2022}. Our approach overcomes the limitations of previous methods that function only at elevated temperatures, leaving pseudomagnetic phenomena largely unexplored~\cite{kovalchuk_non-uniform_nodate,kovalchuk2022non}. Moreover, our clean samples ensure a long lifetime and low decoherence rate of excitons. We focus on two materials representative of the TMDs family: MoSe$_2$, chosen for its well-understood excitonic spectrum~\cite{liu_exciton-polaron_2021}, and WSe$_2$, selected for its long coherence time of excitons comparable to their lifetime~\cite{hao_direct_2016, dufferwiel_valley_2018, boule_coherent_2020, jakubczyk_radiatively_2016}. 

Our device consists of a TMD monolayer suspended over a trench in an Au/SiO$_2$/Si stack (Fig.~\ref{fig:1}d,e). A gate voltage, $V_G$, applied between the Si substrate and the sample induces an electrostatic pressure and strains the TMD, with the strain distribution defined by the trench geometry (see Note~S2 for calibration of applied strain). For an elliptical trench with major axis \(a\) and minor axis \(b\) (\(a \gg b\)), a nearly uniaxial strain is induced along \(b\), which we quantify via the degree of uniaxiality, \(U = (\varepsilon_{bb} - \varepsilon_{aa})/(\varepsilon_{bb} + \varepsilon_{aa})\). Specifically, we use an ellipse with $a = 8$~\(\mu\)m and $b = 3$~\(\mu\)m, which ensures high degree of uniaxiality \(U \approx 80\%\) (Fig.~\ref{fig:1}f), while maintaining strain uniformity \(\frac{\Delta\varepsilon}{\varepsilon} < 10\%\) within the laser spot of $\sim1$~\(\mu\)m (Fig.~S1a-c). Conversely, a device with a circular trench experiences uniform biaxial strain \((U \approx 0)\) in the center of the membrane (Fig.~S1).

In a prototypical experiment, the uniaxial strain generates a pseudomagnetic field, \(\bm{\Omega}\), along the \(x\)-axis in pseudospin space (Fig.~\ref{fig:1}c). In analogy to the Zeeman effect, we expect the exciton energy to depend on the orientation of its pseudospin \(\bm{S}\) with respect to \(\bm{\Omega}\), being minimal when the two vectors are aligned. To study this effect, we use the fact that the pseudospin orientation on the Bloch sphere determines the polarization of a photon coupled to this pseudospin. Specifically, we access the energy of the states with pseudospin along the equator of the Bloch sphere by recording the linear polarization-resolved photoluminescence (PL) spectra.

In analogy to the Larmor effect, the pseudospin along the \(y\)-axis in pseudospin space~---~that is, excited by light polarized along a direction \(45^{\circ}\) with respect to the strain axis~---~undergoes damped precession around \(\bm{\Omega}\) (red cloud in Fig.~\ref{fig:1}c). Such precession is signaled by the appearance of the pseudospin component \(S_z\), while the damped nature of the precession leads to the development of a pseudospin component aligned with the field, \(S_{\parallel}\). We experimentally determine the components of pseudospin from polarization-sensitive PL spectra as \(S_{z} = \frac{ I( \sigma^{+} ) - I(\sigma^{-})}{I( \sigma^{+} ) + I(\sigma^{-})}\) and \(S_{\parallel} = \frac{ I(a) - I(b)}{I(a) + I(b)}\), where \(I (\sigma^{+})\) and \(I(\sigma^{-})\) are the intensities of \(\sigma^{+}\) or \(\sigma^{-}\) polarized light, and $I(a)$ and $I(b)$ are intensities polarized along or perpendicular to the strain axis, respectively~\cite{schmidt_magnetic-field-induced_2016}.

We begin by studying an analog of the Zeeman effect to characterize the field we can achieve. Subsequently, we investigate the Larmor effect in this field. The characteristic time scales extracted from these measurements provide insights into the mechanisms of pseudospin polarization loss and strategy to suppress it. We finally develop a counterpart of $g$-factor measurements to uncover nature of many-body states.

\begin{figure*}[ht]
  \includegraphics[width=1\linewidth]{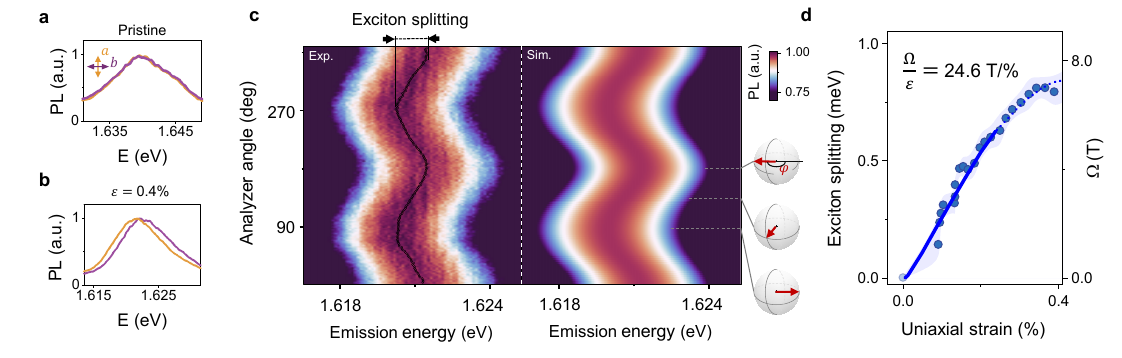}    
  \caption{
    \textbf{Pseudo-Zeeman effect. a,b)} Polarization-resolved PL spectra at near-zero strain (top panel) and under 0.4\% uniaxial strain in the region of neutral exciton X$^0$ in MoSe$_2$. The emission energy of X$^0$ becomes polarization-dependent under strain, with higher energy along the direction of uniaxial strain $b$ (purple) than orthogonal to it (orange) 
    \textbf{c)} Normalized PL spectra for the same device as a function of the analyzer angle at 0.4\% strain, along with the simulations (circles mark the extracted peak position). Note, that the angle \(\varphi\) between the probed pseudospin \(\bm{S}\) and \(\bm{\Omega}\) is twice the angle between the polarizer (analyzer) axis and the strain direction \(b\) (side panel). 
    \textbf{d)} The splitting between PL energy in different polarizations, interpreted as pseudo-Zeeman splitting, extracted from c). The shaded area represents the uncertainty.
  }
  \label{fig:2}
\end{figure*}

\noindent\textbf{Zeeman splitting in pseudomagnetic field.} 
Figure 2a shows the polarization-resolved PL spectra of X$^0$ emission energy in an unstrained MoSe$_2$ (Methods). The orange and purple spectra, corresponding to the polarization along the major (\(a\)) and minor (\(b\)) axes, respectively, show the expected nearly identical emission energy, \(E_{a} = E_{b}\). However, a relative energy shift emerges when an uniaxial strain is applied (\(\varepsilon = \varepsilon_{bb} - \varepsilon_{aa} = 0.4\%\); Fig.~\ref{fig:2}b). Indeed, a false-color map of the polarization-resolved PL spectra of the strained sample (left panel in Fig.~\ref{fig:2}c) reveals a clear sinusoidal dependence of the X$^0$ emission energy on the detection polarization direction. The minimum and maximum of the X$^0$ emission energy correspond to \(\bm{S}\) oriented along and opposite to \(\bm{\Omega}\), respectively (see schematic in Fig.~\ref{fig:2}c). This strain-induced energy splitting between the two orthogonal polarization directions is, in fact, analogous to the Zeeman effect for pseudospins; hence, we term it pseudo-Zeeman splitting.

To quantify the established pseudo-Zeeman effect, we fit the data in Fig.~\ref{fig:2}c using \(E(\varphi) = E_{0} + (\hbar\Omega)/2\,\cos{\varphi}\), where the term \(E_{0} = A(\varepsilon_{xx} + \varepsilon_{yy})/2\) describes the strain-induced redshift in X$^0$ energy compared to the unstrained state (see Eq.~1) and $\varphi$ is the angle between the exciton pseudospin and pseudomagnetic field. The extracted pseudomagnetic field grows linearly at small strain values ($<0.4$\%) at a rate of \(B = 24.6 \pm 2.5\)~T/\% in MoSe$_2$ (solid line in Fig.~\ref{fig:2}d) and \(16.1 \pm 1.8\)~T/\% in WSe$_2$ (Fig.~S2). We used the electron's gyromagnetic factor of \(2\mu_{B} = 0.116~\text{meV/T}\) to convert the units of pseudomagnetic field to Tesla (\(\mu_{B}\) is the Bohr magneton). At higher strain values, the apparent dependence of pseudomagnetic field becomes sublinear (Fig.~S3), which we attribute to a reduced intensity of the higher pseudo-Zeeman-split state when the splitting exceeds the thermal energy ($k_BT \approx 1$~meV). The model based on this mechanism closely aligns with the observed behavior of X$^0$ (simulation in Fig.~\ref{fig:2}c, Note~S3) and extracted splitting (dashed line in Fig.~\ref{fig:2}d), offering further support for this interpretation. Therefore, in the following, we assume a linear dependence of \(\Omega\) on strain, with \(\Omega\) reaching \(42.6 \pm 6.0\)~T in MoSe$_2$ at our highest applied strain of 1.6\% (Fig.~S3). Finally, we note that the pseudo-Zeeman effect is absent in biaxially strained devices (\({\Omega} = 0\)), an experimental situation realized in circular trenches (Fig.~S4). This finding further confirms that the observed dependence in Fig.~\ref{fig:2} results from the pseudospin Zeeman effect and rules out artifacts related to, e.g., spurious plasmonic effects, biaxial strain, etc.

\begin{figure*}
  \includegraphics[width=1\linewidth]{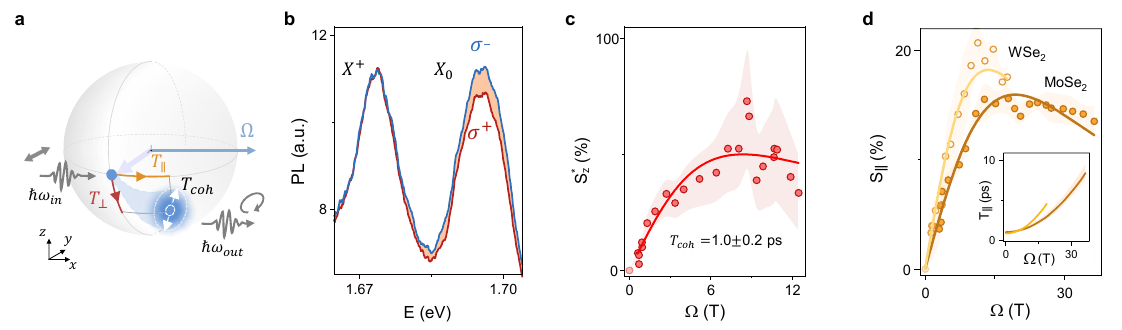}    
  \caption{
    \textbf{Pseudo-Larmor effect. a)} Schematics of the expected Larmor-like dynamics of pseudospins in pseudomagnetic field. 
    \textbf{b)} Circular-polarization-resolved PL spectra of WSe$_2$ under 8~T pseudomagnetic field, under linearly polarized excitation. The rotation of the exciton's pseudospin is manifested as an asymmetry between \(\sigma^{-}\) and \(\sigma^{+}\) emission of the neutral exciton (X$^0$). 
    \textbf{c)} The $S_z^*$ component of the pseudospin vs. the pseudomagnetic field strength in WSe$_2$ (red points) and fit to the model Eq.~\eqref{eq:2} (red line). The shadow represents uncertainty. 
    \textbf{d)} The component of the pseudospin along the field, $S_\parallel$, vs. field strength in MoSe$_2$ and WSe$_2$ and fit to our theoretical model Eq.~\eqref{eq:2}. Inset: the dependence of \(T_{\parallel}\) on the pseudomagnetic field strength in MoSe$_2$ and WSe$_2$ (dark and bright orange lines, respectively).
  }
  \label{fig:3}
\end{figure*}

\noindent\textbf{Strain control of pseudospin dynamics.} Our next objective is to investigate the pseudospin analog of Larmor precession and measure the characteristic pseudospin relaxation times. A hallmark of Larmor precession is the emergence of circularly polarized PL emission under linearly polarized excitation (Fig.~\ref{fig:3}a).

Figure~\ref{fig:3}b shows circular polarization-resolved PL spectra of WSe$_2$ at \(\Omega = 8\)~T (\(\varepsilon = 0.5\%\)). Under this field a prominent asymmetry between the \(I({\sigma}^{+})\) and \(I({\sigma}^{-})\) intensities at the X$^0$ emission energy emerges whose sign depends on the excitation polarization direction (Fig.~S5). This observation is striking, as a circularly polarized emission under linear excitation can only be caused by the breaking of either time-reversal or spatial symmetries. Since the magnetic field is absent in our experiments and an asymmetry is detected only when a pseudomagnetic field is induced (Fig.~S6), it confirms that the pseudomagnetic field alone is responsible for the observed Larmor-like effect.

To gain insight into the mechanism of pseudospin dynamics and relaxation, we develop a theory of pseudo-Larmor precession. The full  model is provided in Note~S1, we illustrate the concept here with an example based on the Bloch equation for population-averaged pseudospin dynamics
\begin{equation} \label{eq:2}
 \frac{\partial\bm{S}}{\partial t} + \frac{\bm{S}}{\tau} + \bm{S}_\perp \times \bm{\Omega} + \frac{\bm{S}_\perp}{T_{coh}} + \frac{\bm{S}_{\parallel} - \bm{S}_{0}}{T_{\parallel}} = \bm{G},
\end{equation}
where \(\bm{G}\) is the pseudospin generation rate defined by the excitation intensity and polarization, \(\bm{S}_{0}\) describes the quasi-equilibrium (thermal) pseudospin induced by the pseudomagnetic field. The characteristic times are (Fig.~\ref{fig:3}a): \(\tau \approx 2\)~ps is the lifetime of an exciton \cite{boule_coherent_2020, dufferwiel_valley_2018, madeo_directly_2020-1, bangeUltrafastDynamicsBright2023, goddeExcitonTrionDynamics2016, chowPhononassistedOscillatoryExciton2017, wangPolarizationTimeresolvedPhotoluminescence2015, yagodkinProbingFormationDark2023b}, \(T_\perp = 2\pi/\Omega\)  is the period of Larmor precession, \(T_{coh}\) is the coherence time that determines relaxation of the pseudospin components transverse to the field, and \(T_{\parallel}\) characterizes the time over which thermal equilibrium between the split sublevels is established (for the relation of Eq.~\eqref{eq:2} to the microscopic model see Notes S1, S4, and S5).
The microscopic model accounts for the exciton longitudinal-transverse splitting caused by the electron-hole exchange interaction. This splitting induces an effective wavevector-dependent pseudomagnetic field $\Omega^{\rm LT}$, which leads to loss of pseudospin coherence by the Dyakonov–Perel mechanism~\cite{glazov_exciton_2014, glazov_coherent_2022}. A strain-induced pseudomagnetic field suppresses  $\Omega^{\rm LT}$-induced depolarization which significantly increases both $T_{coh}$ and $T_\parallel$ (Note~S4). Our goal is to experimentally determine these two timescales that define pseudospin dynamics yet remain unknown.

In a simple case of unitary excitation along the \(y\) pseudospin axis, \(\bm{G}\tau_{\perp} = (0, 1, 0)\), the steady-state solution of Eq.~\eqref{eq:2} is \(S_{z} = {\tau_{\perp}\Omega}/[1 + \left( \tau_{\perp}\Omega \right)^{2}]\), where $1 / \tau_\perp  = 1/T_{coh} + 1/\tau$
(Note~S1). Intuitively, \(S_{z}\) grows linearly with \(\Omega\) when the average rotation angle for pseudospins during their lifetime, \(\tau_{\perp}\Omega\), is smaller than \(2\pi\). At higher field strengths, the pseudospin undergoes multiple rotations around the Bloch sphere during the exciton lifetime, reducing the average pseudospin polarization similarly to the Hanle effect in real magnetic fields. To experimentally realize the scenario of unitary excitation, we consider the reduced pseudospin \(S_{z}^{*}(\Omega)\), normalized to the measured generation rate at the corresponding field \(G(\Omega)\) (Note~S4).


Figure~\ref{fig:3}c shows the experimentally obtained dependence of \(S_{z}^{*}\) on the pseudomagnetic field in WSe$_2$, along with a fit using the solution of Eq.~\eqref{eq:2}. This fit yields \(T_{coh} =\tau_\perp \tau/(\tau - \tau_\perp)= 1.0 \pm 0.2\)~ps in the regime of high field strength, which is longer than the coherence time measured in the unstrained samples (\(T_{coh} \sim 0.5\)~ps~\cite{dufferwiel_valley_2018, boule_coherent_2020}) due to the influence of the pseudomagnetic field (Note~S4). We also note that the average rotation for the pseudospins reaches \(\tau_{\perp}\Omega \approx 6\pi\) at the largest induced \(\Omega\) in WSe$_2$, \(\sim25\)~T (Fig.~S2). Finally, a large polarization, \(S_{z}^{*} = 50\%\), demonstrates the strong potential of the pseudomagnetic field for control of exciton valley pseudospin.

\begin{figure*}[ht]
  \includegraphics[width=1\linewidth]{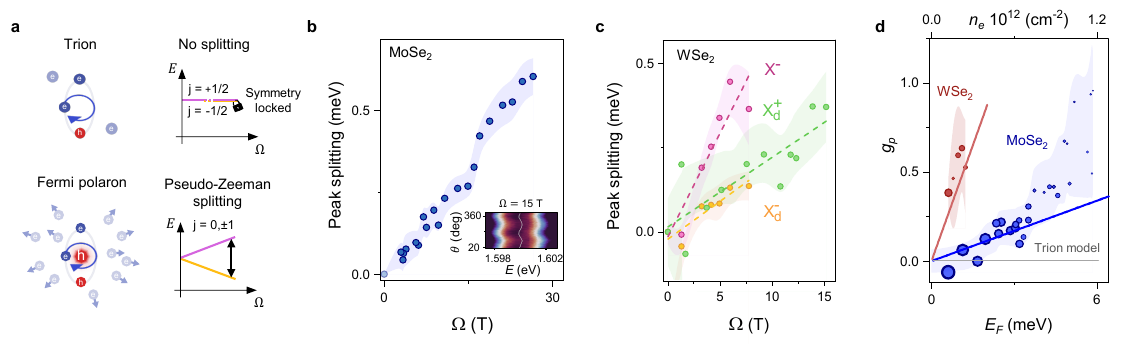}    
  \caption{
    \textbf{Charged excitons under pseudomagnetic field. a)} Schematic illustration of trionic and polaronic representations of charged excitons. \textbf{b)} Splitting of the negatively charged exciton as a function of pseudomagnetic field strength in doped MoSe$_2$. The observed splitting is consistent with the polaronic character of the charged exciton. Inset: False-color map of polarization-resolved PL spectra of the charged exciton in MoSe$_2$. \textbf{c)} Peak splitting of bright (X$^-$) and dark (X$_d^+$, X$_d^-$) charged excitons in WSe$_2$ as a function of pseudomagnetic field strength. \textbf{d)} The dependence of pseudospin $g$-factor \(g_{p}\) of bright FP on Fermi energy in WSe$_2$ (red points) and MoSe$_2$ (blue points), alongside theoretical predictions~\cite{iakovlev_fermi_2023} (red and blue solid lines, respectively). The size of each point is proportional to strain.
  }
  \label{fig:4}
\end{figure*}

To determine \(T_{\parallel}\), we examine Eq.~\eqref{eq:2} under unpolarized excitation conditions, which are experimentally realized at high detuning of the excitation energy from the X$^0$ resonance so light-induced polarization vanishes. In this case $G\to 0$ and only field-induced $\bm S$ appears in the form \(S_{\parallel} = \tau/(\tau + T_{\parallel})\times\tanh\left[ {\hbar\Omega}/({2k_{B}T}) \right]\).
%
%
This expression suggests that the initially unpolarized pseudospins tend to align along \(\bm{\Omega}\), acquiring a pseudospin polarization within a thermal distribution. The induced pseudospin polarization saturates when the pseudo-Zeeman splitting exceeds the thermal energy ($k_BT \approx 1$~meV), with its maximum value determined by the ratio of the relaxation time \(T_{\parallel}\) to the lifetime \(\tau\).

The experimentally observed \(S_{\parallel}\ \text{vs.}\ \Omega\) (Fig.~\ref{fig:3}d) matches these expectations. At low field strengths (\(\Omega < 10\)~T), we observe a linear increase in \(S_{\parallel}\). At higher fields, the polarization reaches the expected plateau, \(S_{\parallel}\left( \hbar\Omega \gg k_{B}T \right) = {\tau}/{(\tau + T_{\parallel})}\). From the value of \(S_\parallel \approx 20\%\) at the plateau in both MoSe$_2$ and WSe$_2$, we find the pseudospin relaxation time \(T_{\parallel} \sim 10\)~ps (Note~S4), significantly longer than the exciton coherence \(T_{coh} \sim 0.5\)~ps and even lifetime \(\tau \approx 2\)~ps in these samples~\cite{robert_exciton_2016, raiber_ultrafast_2022}. This extension arises because the pseudomagnetic field suppresses pseudospin decay dominated by \(\Omega^{\text{LT}}\) (see Note~S4). Using a model that accounts for this effect, we fit \(S_{\parallel}\) and find that the relaxation time increases from 1 to 8~ps over the studied range of field strengths (inset in Fig.~\ref{fig:3}d). Furthermore, this analysis allows us to extract the field responsible for loss of pseudospin coherence, yielding the root-mean-square values \(\Omega_{\text{WSe}_2}^{\text{LT}} = 10.4 \pm 1.3\)~T in WSe$_2$ and \(\Omega_{\text{MoSe}_2}^{\text{LT}} = 12.0 \pm 1.1\)~T in MoSe$_2$ in reasonable agreement with the model predictions (Note~S5). To the best of our knowledge, this constitutes the first measurement of this fundamental parameter.

\textbf{Many-body states under pseudomagnetic field.} Our ultimate goal is to explore the pseudomagnetic field response of many-body states that are more complex than the neutral exciton and to showcase the unique capacity of the pseudomagnetic field to reveal their intrinsic structure. Among these states, the charged excitons (X$^{+/-}$) are the most intriguing ones, as they are described by two alternative models: (i) the ``trion" picture, a three-particle state composed of a Coulomb-bound electron (hole) and two holes (electrons); and (ii) the ``Fermi-polaron" (FP) or Suris tetron picture, where a neutral exciton is correlated with the electron-hole pair in a Fermi sea (Fig.~\ref{fig:4}a)~\cite{suris_correlation_2003, glazov_optical_2020, sidler_fermi_2017, efimkin_electron-exciton_2021, rana_many-body_2020, rana_many-body_2021}. The two pictures are predicted to converge at small densities of charge carriers, when the effect of the polarized charge is negligible (red cloud in Fig.~\ref{fig:4}a). However, the two descriptions are fundamentally different: trions should exhibit fermionic statistics and time-reversal symmetry breaking only under real magnetic fields  due to their odd particle count, while FPs, composed of an even number of particles, are bosons and have linearly polarized states invariant with respect to time-reversal symmetry. Despite their different statistics, no physical phenomena so far, to the best of our knowledge, could conclusively distinguish trions and FPs, leading to a debate about the nature of the observed complexes.

Recent theoretical studies have suggested that the distinct responses of trions and Fermi polarons with respect to time-reversal symmetry may lead to contrasting behaviors under a pseudomagnetic field~\cite{iakovlev_fermi_2023, iakovlev_longitudinal-transverse_2024}. An energy splitting of trions in a pseudomagnetic field is prohibited by time-reversal symmetry due to their fermionic nature, unlike excitons and Fermi polarons, which follow bosonic statistics (Fig.~\ref{fig:4}a). This, in turn, suggests a possibility of a $g$-factor-like measurement in a pseudomagnetic field, \(\Delta E = g_{p}\hbar\Omega\), where \(g_{p}\) is the pseudomagnetic $g$-factor, with \(g_{p} = 0\) signifying a trion and \(g_{p} \neq 0\) indicating a Fermi polaron nature of the charged exciton.

To test these predictions, we probed the response of charged excitons in MoSe$_2$ and WSe$_2$ under a pseudomagnetic field (Fig.~\ref{fig:4}b and Fig.~\ref{fig:4}c). We used the same experimental configuration and analysis as in the study of the pseudo-Zeeman effect of neutral excitons. Figure~\ref{fig:4}b shows the pseudomagnetic-field-induced energy splitting between the negatively charged excitons (X$^{-}$) in doped MoSe$_2$ (\(n_{e} > 1 \times 10^{12}\) cm$^{-2}$) with pseudospin aligned along and opposite to the pseudomagnetic field; the inset shows the false-color map of PL emission vs. polarization angle. A finite energy splitting for X$^{-}$ is similar to what was seen previously for neutral excitons (Fig.~\ref{fig:2}d), although with a much lower amplitude. The observation of pseudo-Zeeman splitting of the charged excitons provides conclusive evidence of their Fermi polaron nature and establishes their bosonic statistics.

In contrast to MoSe$_2$, a monolayer WSe$_2$ hosts a plethora of additional many-body states (Fig.~S2), including positively and negatively charged bright excitons ($\text{X}^{+}$ and $\text{X}^{-}$), neutral and charged dark excitons ($\text{X}_d$, $\text{X}_d^{+}$, and $\text{X}_d^{-}$), biexcitons (XX), and phonon replicas ($\text{X}_p$)~\cite{rivera_intrinsic_2021-1,heValleyPhononsExciton2020}. We observe a considerable energy splitting of $\text{X}^{-}$, $\text{X}_d^{+}$, and $\text{X}_d^{-}$ (Fig.~\ref{fig:4}c), which confirms their Fermi polaronic nature. Interestingly, the dark species demonstrate lower splitting and an overall lower pseudomagnetic $g$-factor, \(g_{p} (\text{X}_d^{+/-}) \approx 0.2\), compared to the bright ones, \(g_{p} (\text{X}^{-}) \approx 0.5\) for the same doping level. The low PL intensity of biexcitons and phonon replicas prevents us from extracting their splitting, while X$^{+}$ is only visible at low pseudomagnetic fields (Fig.~S2).

Finally, we use the pseudomagnetic $g$-factor to explore the effect of Fermi energy (charge density) on the character of charged excitons. Indeed, the polarized charge that distinguishes trions and FPs is strongly affected by Fermi energy, which is reflected in the value of \(g_{p}\). In our measurements, the induced strain is varied together with the Fermi energy; nevertheless, since the pseudomagnetic field is related to the pseudo-Zeeman splitting of neutral excitons ($\Delta E_X = \hbar \Omega$) and the splitting of FP is given by $\Delta E_\text{FP} = g_p\hbar\Omega$, then $g_p = \Delta E_\text{FP}/\Delta E_X$. The pseudomagnetic $g$-factor of FP vs.\ Fermi energy is plotted in Fig.~\ref{fig:4}d; the size of each point is proportional to the uniaxial strain (see Note~S6 for Fermi energy estimation). We find that for low Fermi energy, \(g_{p}\) is nearly zero despite a large pseudomagnetic field, which is consistent with the convergence of Fermi polaronic and trionic pictures in this regime. Meanwhile, at larger $E_F$, the splitting of the charged exciton approaches that of a neutral exciton. This behavior is expected, as the additional charge of the FP becomes effectively screened at high carrier densities, making it similar to a neutral exciton. Indeed, theory predicts~\cite{iakovlev_fermi_2023} that the $g$-factor depends linearly on Fermi energy \(E_{F}\) (Note~S7). Moreover, the predicted value of $g_{p}$ for charged excitons in WSe$_2$ (red line in Fig.~\ref{fig:4}d) is higher than that in MoSe$_2$ (blue line in Fig.~\ref{fig:4}d) for the same doping level, due to their intervalley nature~\cite{iakovlev_fermi_2023}. A close match between the experimental results and theoretical predictions further supports the tuning of FP character by induced charge density. Overall, our results establish the pseudo-Zeeman splitting as a tool to assess the symmetry and statistics of excitonic states.
%
%
\begin{center}
  \textbf{Discussion}
\end{center} 
Our technique to study and manipulate pseudospin opens multiple new possibilities. First, the interplay between magnetic and pseudomagnetic fields in the same device is promising to reveal unique effects~\cite{chen2020theory}. The presence of strongly coupled spin and valley pseudospin degrees of freedom with distinctive timescales should cause complex and hitherto unstudied dynamics. Second, our results indicate rotation of pseudospin by up to \(6\pi\) during pseudo-Larmor precession. This dynamics of pseudospin can be probed in the time domain by observing an oscillating signal in, e.g., time-resolved Kerr rotation microscopy~\cite{sim_ultrafast_2018,raiber_ultrafast_2022}. Third, the coupling between pseudospin and momentum can lead to the pseudomagnetic counterparts of spin-orbit phenomena such as the anomalous Hall, quantum spin Hall, and Rashba-like effects~\cite{bercioux2015quantum,plotnik2016analogue,chen2020theory,rongPhotonicRashbaEffect2020}. The complex nature of momentum/pseudospin coupling should significantly alter these effects compared to their classical counterparts~\cite{barsukova_direct_2024,barczykObservationLandauLevels2024,PhysRevB.94.020301}. Moreover, the high strength of the pseudomagnetic field allows the simulation of magnetic phenomena in the regime of ultra-strong fields (\textgreater100~T) inaccessible to current technologies~\cite{dean2013hofstadter}. Finally, the effects studied above suggest several potential applications. For example, the Larmor precession of pseudospin should generate THz emission with the frequency controlled by the amount of strain, potentially a broadly tunable THz emission source~\cite{yagodkin_ultrafast_2020, ma_recording_2019}. If the coherence time could be extended, e.g. in TMD heterostructures~\cite{zhai2020layer, yagodkinProbingFormationDark2023b}, pseudospin-based devices could be considered as qubits potentially suitable for the effective transduction of mechanical and optical information.
\begin{center}
  \textbf{Methods}
\end{center} 
\textit{Sample fabrication} The devices were fabricated by dry transfer of mechanically exfoliated TMD flakes onto elliptical (8\(\times\)3~\(\mu\)m) or circular trenches (diameter \(\sim\)5~\(\mu\)m), which were wet-etched via hydrofluoric (HF) acid in an Au/Cr/SiO$_2$/Si stack~\cite{hernandezlopezStrainControlHybridization2022,kumar_strain_2024}. The strain in the membrane was induced by applying a gate voltage (typically up to \(\pm\)210~V) between the TMD flake (electrically grounded) and the Si back gate of the chip. The strain in the center was characterized using laser interferometry (see Note~S2).

\textit{Optical measurements}
The devices were measured inside a cryostat (CryoVac Konti Micro) at a base temperature of 10~K. Photoluminescence (PL) measurements were carried out using a Kymera 193i spectrograph and continuous-wave (CW) lasers with either \(\lambda = 685\)~nm (8~\(\mu\)W) for quasi-resonant excitation or \(\lambda = 532\)~nm (6~\(\mu\)W) for detuned excitation. The lasers were tightly focused at the center of the membrane with a spot diameter of approximately 0.8~\(\mu\)m. The excitation polarization was controlled using a half-wave plate (RAC 4.2.10, B. Halle) placed before the objective (Olympus LMPlan 50x, 0.5~NA) to reduce polarization loss. The detection polarization was set using a combination of either a half-wave plate or a quarter-wave plate (for linear and circular detection, respectively) and an analyzer (GL~10, Thorlabs) before the spectrometer. To minimize the influence of coherent effects on pseudo-Zeeman splitting, we maintained excitation and detection co-polarized. The Fermi polaron splitting was measured in a Cryostation s100 cryostat (Montana Instruments) with an Isoplane 320 spectrometer (Teledyne Princeton Instruments), using a 532~nm CW laser focused to a diffraction-limited spot with an objective (Zeiss Epiplan 100x, 0.75~NA).

\textbf{Acknowledgements}
The Berlin groups acknowledge the Deutsche Forschungsgemeinschaft (DFG) for financial support through the Collaborative Research Center TRR 227 Ultrafast Spin Dynamics (project B08) and the Priority Programme SPP 2244 2DMP (project BO 5142/5) and the Federal Ministry of Education and Research (BMBF, Projekt 05K22KE3). The Saint Petersburg group acknowledges financial support by the RSF Project  23-12-00142 (theory); Z.A.I gratefully acknowledges
the BASIS foundation. K.I.B. acknowledges illuminating discussions with Christiane Koch.

\textbf{Author Contributions}
D.Y. and K.I.B. conceived the project. Z.A.I and M.M.G. developed the theory. D.Y., A.M.K., A.D., and C.G. designed the experimental setup. D.Y., K.B., A.M.K., and B.H. prepared the samples. D.Y., K.B., A.M.K., and A.D. performed the optical measurements. O.Y. performed mechanical simulations. D.Y. and K.B. analyzed the data. D.Y., and K.I.B. wrote the manuscript with input from all co-authors.

\textbf{Data Availability Statement}
The data that support the findings of this study are available from the
corresponding author upon reasonable request.

The authors declare no competing interest.

\bibliographystyle{naturemag}
\bibliography{supplementary}

\ifarXiv
    

\section*{Supplementary material}

# S1 Theory of pseudospin in pseudomagnetic field

## S1.1 Pseudospin relaxation

We describe exciton pseudospin dynamics within the framework of the kinetic equation for the $2 \times 2$ density matrix [1, 2, 3]

$$\rho_k = n_k \hat{\sigma}_0 + (s_k \cdot \hat{\sigma}).$$

Here $k$ is the exciton in plane wavevector, $n_k = \mathrm{Sp}\{\rho_k\}/2$ is the pseudospin averaged occupancy of the orbital state with the wavevector $k$, $s_k = \mathrm{Sp}\{\rho_k \hat{\sigma}\}/2$ is the average pseudospin in this orbital state; $\hat{\sigma}_0$ is the unit $2 \times 2$ matrix and $\hat{\sigma} = (\hat{\sigma}_x, \hat{\sigma}_y, \hat{\sigma}_z)$ is the vector composed of the Pauli matrices. The kinetic equation for the vector $s_k$ reads (the time argument is omitted for brevity)

$$\frac{\partial s_k}{\partial t} + s_k \times \Omega_k + \frac{s_k}{\tau} + \mathrm{St}\{s_k\} = g_k, \tag{S1}$$

where $\tau$ is the exciton lifetime, $g_k$ is the pseudospin generation rate (assumed to be independent of orientation of $k$),

$$\mathrm{St}\{s_k\} = \frac{s_k - \langle s_k \rangle}{\tau_{sc}}, \quad \langle s_k \rangle = \int_0^{2\pi} \frac{d\varphi_k}{2\pi} s_k,$$

is the collision integral described, for simplicity, by a single relaxation time $\tau_{sc}$ responsible for isotropization of the non-equilibrium distribution function, $\varphi_k$ is the polar angle of the wavevector $k$, and $\Omega_k$ is the effective pseudospin precession frequency (pseudomagnetic field) that consists of two contributions

$$\Omega_k = \Omega + \Omega_k^{LT}, \tag{S2}$$

with $\Omega \parallel x$ being the momentum-independent contribution related to the anisotropic strain and $\Omega_k^{LT}$ is the exciton longitudinal-transverse splitting. The latter can be recast as

$$\Omega_k^{LT} = \Omega_k^{LT} \times (\cos 2\varphi_k, \sin 2\varphi_k, 0),$$

with $\Omega_k^{LT} \propto k$, see Refs. [4, 5, 6, 3] for details. Equation (S1) disregards the (slow) energy relaxation processes, they are taken into account in the next subsection S1.2.

Under assumptions of $\Omega_k^{LT} \tau_{sc} \ll 1$ and $\tau_{sc} \ll \tau$ Eq. (S1) can be solved by iterations in the small parameter $\Omega_k^{LT} \tau_{sc}$ with the resulting equation for the angular average pseudospin, $\langle s_k \rangle$, in the form

$$-\mathrm{i}\omega \langle s_k \rangle_\omega + [1 - \Xi(\omega)]\langle s_k \rangle_\omega \times \Omega + \frac{\langle s_k \rangle_\omega}{\tau} + \hat{\Gamma}(\omega)\langle s_k \rangle_\omega = g_{k,\omega}. \tag{S3}$$

Here the subscript $\omega$ describes the corresponding Fourier component

$$\langle s_k \rangle_\omega = \int_{-\infty}^\infty dt e^{\mathrm{i}\omega t} \langle s_k(t) \rangle, \quad g_{k,\omega} = \int_{-\infty}^\infty dt e^{\mathrm{i}\omega t} g_k(t),$$

and $\hat{\Gamma}(\omega)$ is the effective relaxation rate tensor, and $\Xi(\omega)\Omega$ is effective renormalization of pseudomagnetic field. The components of the tensor $\hat{\Gamma}(\omega)$ and scalar $\Xi(\omega)$ depend on the frequency and are, in general, complex. Introducing

$$\tau_\omega = \frac{\tau_{sc}}{1 - \mathrm{i}\omega\tau_{sc}},$$

we express these parameters as
$$\Xi(\omega) = \frac{(\Omega_k^{LT}\tau_\omega)^2}{2[1+(\Omega\tau_\omega)^2]}, \tag{S4}$$

and
$$\Gamma_{xx}(\omega) = \Gamma_{yy}(\omega) = \frac{(\Omega_k^{LT})^2 \tau_\omega}{2[1+(\Omega\tau_\omega)^2]}, \tag{S5a}$$

$$\Gamma_{zz}(\omega) = \Gamma_{xx}(\omega) + \frac{(\Omega_k^{LT})^2 \tau_\omega}{2}, \tag{S5b}$$

Equation (S3) is the analogue of the pseudospin Bloch equation [Eq. (2) of the main text] microscopically derived for the relevant situation where the pseudospin relaxation and dynamics is controlled by the longitudinal-transverse splitting of excitons $\Omega_k^{LT}$. The principal issue here is that both effective pseudospin precession frequency $\Omega[1 - \Xi(\omega)]$ and the relaxation rates $\hat{\Gamma}(\omega)$ are frequency dependent. It provides a memory effect in pseudospin dynamics and makes it, generally, more complicated compared to that described by the classical Bloch equation. For example, the expressions for $\Gamma_{xx}$, $\Gamma_{yy}$, and $\Gamma_{zz}$ at $\omega = \Omega$ are presented in Ref. [3], Eqs. (10); while the experimentally relevant case of $\omega = 0$ is studied below.

In accordance with experimental situation we focus on the steady-state excitation where $g_k = \text{const}(t)$. Setting $\omega = 0$ in Eqs. (S5) renders both the scalar $\Xi(0)$ and the tensor $\hat{\Gamma}(0)$ real [cf. Ref. [7]] and we obtain

$$\Gamma_{xx} = \frac{\Gamma_s}{1+(\Omega\tau_{sc})^2}, \tag{S6a}$$

$$\Gamma_{yy} = \frac{\Gamma_s}{1+(\Omega\tau_{sc})^2}, \tag{S6b}$$

$$\Gamma_{zz} = \frac{\Gamma_s}{1+(\Omega\tau_{sc})^2} + \Gamma_s, \tag{S6c}$$

where
$$\Gamma_s = \frac{(\Omega_k^{LT})^2 \tau_{sc}}{2}. \tag{S7}$$

The pseudomagnetic field correction $\Xi(0)$ can be neglected for steady-state excitation.

Equations above were derived at $\Omega_k^{LT}\tau_{sc} \ll 1$. At arbitrary value of the product $\Omega_k^{LT}\tau_{sc}$ we obtain at $\omega = 0$

$$\Gamma_{xx} = \Gamma_{yy} = \frac{(\Omega_k^{LT})^2 \tau_{sc}}{1+(\Omega\tau_{sc})^2 + \sqrt{\left[1+(\Omega\tau_{sc})^2 + (\Omega_k^{LT}\tau_{sc})^2\right]^2 - (2\Omega\Omega_k^{LT}\tau_{sc}^2)^2}}, \tag{S8a}$$

$$\Gamma_{zz} = \Gamma_{xx} + \frac{\sqrt{\left[1+(\Omega\tau_{sc})^2 + (\Omega_k^{LT}\tau_{sc})^2\right]^2 - (2\Omega\Omega_k^{LT}\tau_{sc}^2)^2}\,(\Omega_k^{LT})^2 \tau_{sc}}{1+(\Omega\tau_{sc})^2 + \sqrt{\left[1+(\Omega\tau_{sc})^2 + (\Omega_k^{LT}\tau_{sc})^2\right]^2 - (2\Omega\Omega_k^{LT}\tau_{sc}^2)^2}}, \tag{S8b}$$

$$\Xi(0) = \Gamma_{xx}\tau_{sc}. \tag{S8c}$$

The steady-state solution of Eq. (S3) can be written in the cases of interest as follows.

4

1. *Linearly polarized excitation along in the principal axes of deformation, $g_k \parallel x$*

$$\langle s_{k,x} \rangle = \frac{g_{k,x}}{\tilde{\Gamma}_{xx}}, \quad \langle s_{k,y} \rangle = \langle s_{k,z} \rangle = 0. \tag{S9a}$$

2. *Linearly polarized excitation at $45°$ to the prinicpal axes of deformation, $g_k \parallel y$*

$$\langle s_k \rangle = \frac{\tilde{\Gamma}_{zz} g_k}{\tilde{\Omega}^2 + \tilde{\Gamma}_{yy}\tilde{\Gamma}_{zz}} + \frac{\tilde{\Omega} \times g_k}{\tilde{\Omega}^2 + \tilde{\Gamma}_{yy}\tilde{\Gamma}_{zz}}. \tag{S9b}$$

3. *Circularly polarized excitation, $g_k \parallel z$*

$$\langle s_k \rangle = \frac{\tilde{\Gamma}_{yy} g_k}{\tilde{\Omega}^2 + \tilde{\Gamma}_{yy}\tilde{\Gamma}_{zz}} + \frac{\tilde{\Omega} \times g_k}{\tilde{\Omega}^2 + \tilde{\Gamma}_{yy}\tilde{\Gamma}_{zz}}. \tag{S9c}$$

Here

$$\tilde{\Omega} = [1 - \Xi(0)]\Omega$$

takes into account the effect of renormalization at $\Omega_k^{LT}\tau_{sc} \gtrsim 1$, the effect of the finite lifetime can be taken into account via renormalization

$$\tilde{\Gamma}_{\alpha\beta} = \Gamma_{\alpha\beta} + \delta_{\alpha\beta}\frac{1}{\tau}, \quad \alpha, \beta = x, y, z,$$

where $\delta_{\alpha\beta}$ is the Kronecker's $\delta$-symbol.

## S1.2 Energy relaxation and pseudomagnetic field induced polarization

In addition to the optically induced pseudospin polarization, the presence of the deformation-induced splitting results in the quasi-equilibrium, thermal, polarization of excitons.

To calculate the thermal quasi-equilibrium polarization of excitons we need to take into account both pseudospin and energy relaxation processes. To simplify the analysis we assume that the isotropization time of the exciton distribution function ($\tau_{sc}$) is much shorter that both lifetime and energy relaxation time and consider only the "slow" processes. We are interested in the spin polarization along the direction of the pseudomagnetic field, thus, instead of the full $2 \times 2$ spin density matrix it is sufficient to consider only the occupancies of the excitons with the pseudospins parallel and antiparallel to the $x$-axis. We denote these states as $\uparrow$ and $\downarrow$ for brevity. The system of kinetic equations for the spin up ($\uparrow$) and spin down ($\downarrow$) excitons reads

$$\frac{f_\uparrow(\varepsilon)}{\tau} + \frac{f_\uparrow(\varepsilon) - f_\downarrow(\varepsilon)}{2\tau_s} + \frac{d}{d\varepsilon}\left[\hat{J}f_\uparrow(\varepsilon)\right] = \delta(\varepsilon - E_0)g_0, \tag{S10a}$$

$$\frac{f_\downarrow(\varepsilon)}{\tau} - \frac{f_\uparrow(\varepsilon) - f_\downarrow(\varepsilon)}{2\tau_s} + \frac{d}{d\varepsilon}\left[\hat{J}f_\downarrow(\varepsilon)\right] = \delta(\varepsilon - E_0)g_0. \tag{S10b}$$

Here, $\varepsilon$ is the exciton energy, $f_\uparrow(\varepsilon), f_\downarrow(\varepsilon)$ are the distribution functions of excitons with spin up and down, respectively; $\tau_s(\varepsilon) = 1/\Gamma_{xx}(\varepsilon)$ is the spin relaxation time along the direction of the magnetic field. Note, that here we consider $\Gamma_{xx}$ (not $\tilde{\Gamma}_{xx}$) as we take into account finite exciton

lifetime separately. The energy relaxation is described as the energy diffusion (cf. Ref. [8]). The energy flux operator is defined as

$$\hat{J} = D(\varepsilon)\left(\frac{1}{k_B T} + \frac{d}{d\varepsilon}\right), \tag{S11}$$

where $D(\varepsilon)$ is energy diffusion coefficient. The right hand sides of Eqs. (S10) correspond to the spin-isotropic pumping at the high energy $E_0 \gg k_B T$. In what follows, for simplicity, we disregard the energy dependence of the spin relaxation time replacing the relaxation time by its typical value $\tau_s(\varepsilon) \equiv \tau_s$.

We assume that the energy spans from $-\hbar\Omega/2$ to $+\infty$. The bottom of the spin up subband is, thus, at $+\hbar\Omega/2$ and the bottom of the spin down subband is at $-\hbar\Omega/2$. In the narrow energy band, $-\hbar\Omega/2 < \varepsilon < \hbar\Omega/2$, there exists only spin down states, and the kinetic equation simplifies to

$$\frac{f_\downarrow(\varepsilon)}{\tau} + \frac{d}{d\varepsilon}\left[\hat{J}f_\downarrow(\varepsilon)\right] = 0, \tag{S12}$$

while $f_\uparrow$ is irrelevant for this energy band. Here, we assume that the spin splitting is small in comparison with the temperature, $\hbar\Omega \ll k_B T$.

The boundary conditions are straightforwardly formulated in terms of fluxes. In particular, there is no current at the bottoms of subbands

$$\hat{J}f_\uparrow\left(\frac{\hbar\Omega}{2}\right) = 0, \qquad \hat{J}f_\downarrow\left(-\frac{\hbar\Omega}{2}\right) = 0. \tag{S13}$$

The exciton generation at $\varepsilon = E_0$ can also be recast as a boundary condition in the form

$$\hat{J}f_\uparrow(E_0) = \hat{J}f_\downarrow(E_0) = -g_0. \tag{S14}$$

It is convenient to rewrite the boundary conditions (S13) in somewhat different form. As we consider $\hbar\Omega \ll k_B T$, the distribution function $f_\downarrow(\varepsilon)$ can be taken as a constant in the range $-\hbar\Omega/2 < \varepsilon < \hbar\Omega/2$. After the integration of Eq. (S12) over energy $\varepsilon$ from $-\hbar\Omega/2$ to $\hbar\Omega/2$, we express the energy flux at $\varepsilon = \hbar\Omega/2$ as

$$\hat{J}f_\downarrow\left(\frac{\hbar\Omega}{2}\right) = -\frac{\hbar\Omega}{\tau}f(0). \tag{S15}$$

Next, we rewrite the kinetic equations in terms of the exciton population distribution function

$$f(\varepsilon) = \frac{f_\uparrow(\varepsilon) + f_\downarrow(\varepsilon)}{2} = \langle n_k \rangle \tag{S16}$$

and the exciton spin distribution function

$$s_x(\varepsilon) = \frac{f_\uparrow(\varepsilon) - f_\downarrow(\varepsilon)}{2} = \langle s_{k,x}\rangle. \tag{S17}$$

The kinetic equations take a form

$$\frac{f}{\tau} + \frac{d}{d\varepsilon}\left[\hat{J}f(\varepsilon)\right] = 0, \tag{S18a}$$

6

$$\frac{s_x}{\tau} + \frac{s_x}{\tau_s} + \frac{d}{d\varepsilon}\left[\hat{J}s_x(\varepsilon)\right] = 0, \tag{S18b}$$

with the boundary conditions

$$\hat{J}f(0) = -\hat{J}s_x(0) = -\frac{\hbar\Omega}{2\tau}\left[f(0) - s_x(0)\right], \tag{S19a}$$

$$\hat{J}f(E_0) = -g_0, \qquad \hat{J}s_x(E_0) = 0. \tag{S19b}$$

Note that for these equations it is already unimportant where the bottom of the band is set provided that $\hbar\Omega \ll k_B T$. Hereafter we set the bottom energy boundary to zero. For the same reason, the spin polarization $s_x(\varepsilon) \propto \hbar\Omega$ is small in comparison with $f(\varepsilon)$ and can be neglected in the boundary condition (S19a). Moreover, we can entirely omit the $\propto \hbar\Omega$ contribution in the boundary condition for $\hat{J}f(0)$ as we are working in the lowest order in pseudomagnetic field. Thus, the boundary conditions simplify to

$$\hat{J}f(0) = 0, \qquad \hat{J}s_x(0) = \frac{\hbar\Omega}{2\tau}f(0), \tag{S20a}$$

$$\hat{J}f(E_0) = -g_0, \qquad \hat{J}s_x(E_0) = 0, \tag{S20b}$$

and the kinetic equations separates: first, we can solve the kinetic equation for population $f(\varepsilon)$, and then substitute the results in boundary conditions for exciton spin $s_x(\varepsilon)$ and solve for $s_x(\varepsilon)$. Finally, we consider the lifetime sufficiently large, $D(\varepsilon)\tau \gg (k_B T)^2$, such that excitons manage to thermalize as

$$f(\varepsilon) = f_0 \exp\left(-\frac{\varepsilon}{k_B T}\right). \tag{S21}$$

Integrating Eq. (S18a) over $\varepsilon$, we get the obvious result, that the exciton generation rate is equal to their annihilation rate

$$-\hat{J}f(E_0) = g_0 = \frac{1}{\tau}\int_0^{E_0} f_0 \exp\left(-\frac{\varepsilon}{k_B T}\right) d\varepsilon = \frac{k_B T}{\tau}f(0). \tag{S22}$$

Similar integration of the kinetic equation (S18b) for spin gives

$$\left(\frac{1}{\tau} + \frac{1}{\tau_s}\right)\int_0^{E_0} s_x(\varepsilon)d\varepsilon = \hat{J}s_x(0) = \frac{\hbar\Omega}{2\tau}f(0). \tag{S23}$$

The total exciton pseudospin contains also the contribution from the ↓ particles with the energies $-\hbar\Omega/2 < \varepsilon < \hbar\Omega/2$:

$$\int_{-\hbar\Omega/2}^{\hbar\Omega/2} s_x(\varepsilon)d\varepsilon = -\frac{1}{2}\int_{-\hbar\Omega/2}^{\hbar\Omega/2} f_\downarrow(\varepsilon)d\varepsilon = -\hbar\Omega\frac{f(0)}{2}. \tag{S24}$$

Combining Eqs. (S23) and (S24) we obtain

$$\int_{-\hbar\Omega/2}^{E_0} s_x(\varepsilon)d\varepsilon = -\frac{\tau}{\tau + \tau_s}\frac{\hbar\Omega}{2}f(0) = -\frac{\tau}{\tau + \tau_s}\frac{\hbar\Omega}{2k_B T}\int_{-\hbar\Omega/2}^{E_0} f(\varepsilon)d\varepsilon. \tag{S25}$$

7

Normalizing exciton pseudospin to exciton occupancy we have finally

$$S_0 = -\frac{\tau}{\tau + \tau_s} \frac{\hbar \Omega}{2k_B T}. \tag{S26}$$

Equation (S26) shows that the pseudomagnetic field induced pseudospin polarization is controlled by the dynamical factor $\tau/(\tau + \tau_s)$ that describes the probability for exciton to flip its pseudospin during the lifetime. In the general case of arbitrary ratio of $\hbar\Omega$ and $k_B T$ we have for the induced spin polarization $S_0$

$$S_0 = -\frac{\tau}{\tau + \tau_s} \tanh\left(\frac{\hbar\Omega}{2k_B T}\right) \frac{\Omega}{\Omega}. \tag{S27}$$

It is a linear function of $\Omega$ for $\hbar\Omega \ll k_B T$ and saturates at $-\tau/(\tau+\tau_s) \times \Omega/\Omega$ at $\hbar\Omega \gg k_B T$ and $\tau_s \ll \tau$.[1]

To summarize the ensemble averaged longitudinal component of the pseudospin takes the form, see Eqs. (S9a) and (S27):

$$S_\parallel = \frac{\tau}{\tau + T_\parallel} \left[T_\parallel G_\parallel - \tanh\left(\frac{\hbar\Omega}{2k_B T}\right) \frac{\Omega}{\Omega}\right]. \tag{S28}$$

where for convenience of comparison with experiments we introduced $T_\parallel = \tau_s$.

It is noteworthy that for thermalized ensemble of excitons, the ensemble averaging of Eqs. (S9) should be carried out. To avoid additional complications we assume that the pseudospin relaxation rates $\Gamma_{\alpha\beta}$ are appropriately averaged in Eqs. (S9).

### S1.3 Photoluminescence spectra of excitons in pseudomagnetic field

To derive compact analytical expressions for the emission spectra we develop a simplified model description of optically active exciton in 2D TMDC based on semiclassical (oscillator) model. Let $\mathbf{p}(t) = [p_x(t), p_y(t)]$ be the dipole moment of the exciton with $k = 0$, i.e., the exciton which directly emits light. The components $p_\alpha(t)$ ($\alpha = x$ or $y$) obey the driven oscillator equations

$$i\dot{p}_\alpha = \omega_\alpha p_\alpha - i\gamma p_\alpha + \xi_\alpha. \tag{S29}$$

Here dot denotes the time derivative, $\omega_\alpha$ is the exciton resonance frequency, $\omega_x \ne \omega_y$ is due to the strain-induced anisotropic splitting, $\gamma$ is the exciton damping (for simplicity we disregard its dependence on $\alpha$) and $\xi_\alpha(t)$ are the random sources related to the exciton relaxation from the high momentum to the optically active states. These random sources are the analogues of Langevin forces in the theory of fluctuations [9, 10]. The random sources approach can be justified for calculation of exciton emission spectra in the case of non-resonant pumping and linear-in-the-pumping regime [11, 12].

Let us calculate the emission spectrum using Eq. (S29). Since the Maxwell equations are linear, the spectrum of emitted electric field is proportional to the spectrum of $\mathbf{p}(t)$. We present

$$p_{\alpha,\omega} = \int_{-\infty}^{\infty} dt e^{i\omega t} p_\alpha(t) = \int_{-\infty}^{\infty} dt \int_{-\infty}^{\infty} dt_1 e^{i\omega t} G_\alpha(t - t_1) \xi_\alpha(t_1), \tag{S30}$$

---
[1] Note that according to Eq. (S6a) $\Gamma_{xx} = \tau_s^{-1}$ depends on $\Omega$. Formally, for very large $\Omega\tau$ the relaxation time $\tau_s$ can exceed $\tau$.

8

where $G_\alpha(t)$ is the Green's function of Eq. (S29). The following correlator reads

$$\langle p_{\alpha,\omega} p^*_{\beta,\omega'} \rangle = \int_{-\infty}^{\infty} dt \int_{-\infty}^{\infty} dt_1 \int_{-\infty}^{\infty} dt' \int_{-\infty}^{\infty} dt_2 e^{i\omega t - i\omega' t'} G_\alpha(t-t_1) G_\beta^*(t'-t_2) \langle \xi_\alpha(t_1) \xi_\beta^*(t_2) \rangle, \quad (S31)$$

where the averaging takes place over the realization of the random sources. In the random sources approach, the correlations of the sources are given by the correlation matrix

$$\langle \xi_\alpha(t_1) \xi_\beta^*(t_2) \rangle = \delta(t_1 - t_2) \Xi_{\alpha\beta}, \quad (S32)$$

while other correlators, e.g., $\langle \xi_\alpha(t_1) \xi_\beta(t_2) \rangle$ are zero. The matrix elements of $\Xi_{\alpha\beta}$ can be related to the exciton generation rates in the density matrix approach, see below. Performing integrations, we find that

$$\langle p_{\alpha,\omega} p^*_{\beta,\omega'} \rangle = 2\pi \delta(\omega - \omega') (p_\alpha p_\beta^*)_\omega, \quad (S33)$$

where the spectrum is

$$(p_\alpha p_\beta^*)_\omega = \int_{-\infty}^{\infty} d\tau \int_{-\infty}^{\infty} dt e^{i\omega\tau} G_\alpha(\tau - t) G_\beta^*(-t) \Xi_{\alpha\beta} = G_{\alpha,\omega} G_{\beta,\omega}^* \Xi_{\alpha\beta}, \quad (S34)$$

with

$$G_{\alpha,\omega} = \frac{1}{\omega - \omega_\alpha + i\gamma}. \quad (S35)$$

Note that if the dynamics of dipoles is described by more complex linear equations with the matrix Hamiltonian $H$ as $i\hbar \dot p = Hp$, the expressions (S34) are generalized, via the matrix Green's function $G_\omega$, as

$$(p_\alpha p_\beta^*)_\omega = \sum_{\alpha'\beta'} G_{\alpha\alpha',\omega} G_{\beta\beta',\omega}^* \Xi_{\alpha'\beta'}, \quad G_{\alpha\beta} = [G]_{\alpha\beta}, \quad G = \frac{\hbar}{\omega I - H}, \quad (S36)$$

where $I$ is the unit $2 \times 2$ matrix.

Now we turn to the calculation of the emission spectra. If we consider the analyzer at the angle $\varphi$, the emission spectra (up to a common factor) can be expressed as

$$I(\varphi) \propto \left( |E_x \cos\varphi + E_y \sin\varphi|^2 \right)_\omega \propto \cos^2\varphi \left( |p_x^2| \right)_\omega + \sin^2\varphi \left( |p_y^2| \right)_\omega + \frac{\sin 2\varphi}{2} \left( p_x p_y^* + p_y p_x^* \right)_\omega. \quad (S37)$$

As a result, by virtue of Eq. (S34) the emission spectra measured in $x$, $y$ and arbitrary polarizations can be expressed via the correlators of the random sources $\Xi_{\alpha\beta}$ as

$$I_x(\omega) = (|p_x|^2)_\omega = \frac{\Xi_{xx}}{(\omega - \omega_x)^2 + \gamma^2}, \quad (S38)$$

$$I_y(\omega) = (|p_y|^2)_\omega = \frac{\Xi_{yy}}{(\omega - \omega_y)^2 + \gamma^2}, \quad (S39)$$

$$I(\varphi) = \cos^2\varphi \frac{\Xi_{xx}}{(\omega - \omega_x)^2 + \gamma^2} + \sin^2\varphi \frac{\Xi_{yy}}{(\omega - \omega_y)^2 + \gamma^2}$$
$$+ \sin 2\varphi \frac{\Xi_{xy}[\gamma^2 + (\omega - \omega_x)(\omega - \omega_y)]}{[(\omega - \omega_x)^2 + \gamma^2][(\omega - \omega_y)^2 + \gamma^2]}. \quad (S40)$$

Here, for simplicity, we assumed that $\Xi_{xy} = \Xi_{yx}$. As we show below, it corresponds to the linearly polarized generation of excitons.

Let us now provide a relation between the random sources correlation matrix $\Xi_{\alpha\beta}$ and exciton generation rates in the density matrix approach which we used previously. To that end, we note that the single time correlators (mean squares, i.e., the Stokes parameters) can be expressed as[2]

$$\langle p_\alpha(t) p_\beta^*(t) \rangle = \int_{-\infty}^{\infty} \frac{(p_\alpha p_\beta^*)_\omega}{2\pi} d\omega. \tag{S41}$$

We obtain

$$\langle |p_\alpha(t)|^2 \rangle = \frac{\Xi_{\alpha\alpha}}{2\gamma}, \tag{S42}$$

$$\langle p_\alpha(t) p_\beta^*(t) \rangle = \frac{\Xi_{\alpha\beta}}{2\gamma + i(\omega_\alpha - \omega_\beta)}. \tag{S43}$$

On the other hand, within the spin density matrix approach the average exciton occupancy

$$N = g\tau = \frac{1}{2} \left( \langle |p_x(t)|^2 \rangle + \langle |p_y(t)|^2 \rangle \right), \tag{S44}$$

where $\tau_{k=0} = 2\gamma$ is the exciton lifetime at $k=0$ state. Hence, the exciton generation rate $g$

$$g = \frac{\Xi_{xx} + \Xi_{yy}}{2}. \tag{S45}$$

At the same time, the pseudospin $S$ components

$$S_x = \left( \langle |p_x(t)|^2 \rangle - \langle |p_y(t)|^2 \rangle \right), \tag{S46a}$$

$$S_y = \left( \langle p_x(t) p_y^*(t) \rangle + \langle p_y(t) p_x^*(t) \rangle \right), \tag{S46b}$$

$$S_z = \left[ \langle p_x(t) p_y^*(t) \rangle - \langle p_y(t) p_x^*(t) \rangle \right], \tag{S46c}$$

obey the steady-state kinetic equation

$$\mathbf{S} \times \mathbf{\Omega} + \frac{\mathbf{S}}{\tau_{k=0}} = \mathbf{g}, \tag{S47}$$

where the vector generation rate $\mathbf{g}$ accounts for both induced and thermal polarization, cf. Notes S1.1 and S1.2.

Comparing the results of the solution of Eq. (S47) with Eqs. (S43), we obtain under linearly polarized excitation ($g_z \equiv 0$ and $\Xi_{xy} = \Xi_{yx}$)

$$S_x = \frac{1}{2}\tau_{k=0}(\Xi_{xx} - \Xi_{yy}) = \tau_{k=0} g_x, \quad S_y = \frac{1}{2}\Xi_{xy}\left(\frac{1}{2\gamma + i\Omega} + \frac{1}{2\gamma - i\Omega}\right) = \frac{g_y \tau_{k=0}}{1 + (\Omega \tau_{k=0})^2}, \tag{S48}$$

hence,

$$g_x = \frac{\Xi_{xx} - \Xi_{yy}}{2}, \quad g_y = \frac{\Xi_{xy} + \Xi_{yx}}{2}, \quad g_z = \frac{i(\Xi_{xy} - \Xi_{yx})}{2}. \tag{S49}$$

---

[2] To that end it is sufficient to compare the definition: $\langle p_\alpha(t) p_\beta^*(t) \rangle = \int dt' G_\alpha(t - t') G_\beta^*(t - t') \Xi_{\alpha\beta}$ with the integral over $\omega$ of Eq. (S34).

10

Finally, we express the emission spectra Eq. (S38) in terms of generation rate Eqs. (S45), (S49) as

$$I_x(\omega) \propto (|p_x|^2)_\omega = \frac{g + g_x}{(\omega - \omega_x)^2 + \gamma^2}, \quad I_y(\omega) \propto (|p_y|^2)_\omega = \frac{g - g_x}{(\omega - \omega_y)^2 + \gamma^2}, \tag{S50}$$

or

$$I(\varphi) \propto N \cos^2 \varphi \frac{g + g_x}{(\omega - \omega_x)^2 + \gamma^2} + N \sin^2 \varphi \frac{g - g_x}{(\omega - \omega_y)^2 + \gamma^2}$$
$$+ N \sin 2\varphi \frac{g_y[\gamma^2 + (\omega - \omega_x)(\omega - \omega_y)]}{[(\omega - \omega_x)^2 + \gamma^2][(\omega - \omega_y)^2 + \gamma^2]}. \tag{S51}$$

We use the resulting dependence of the detected emission spectra on the analyzer angel $\varphi$ to simulate the spectra of pseudo Zeeman split states in the Note S3.

## S2 Calibration of strain

In this note we discuss our approach to define the strain tensor in the suspended monolayer. Frist, we note that the shift of exciton emission energy, $\Delta E$, depends solely on the biaxial strain $(\varepsilon_{xx} + \varepsilon_{yy})/2$ and a material-specific constant $A$ (Eq. (1) in the main text). This constant is determined through calibration measurements using a circular device, whose geometry simplifies the extraction of biaxial strain. With $A$ known, we can estimate the biaxial strain in an arbitrary sample from the energy shift of its exciton emission line. Knowing the biaxial strain allows us to find the uniaxial strain, $\varepsilon_{xx} - \varepsilon_{yy}$, using the uniaxiality factor $U$ defined in the main text. We quantitatively assess the strain in our devices using laser interferometry. The devices are modeled as optical cavities comprising a suspended two-dimensional material flake and the underlying Si/SiO$_2$ substrate (see Fig. S7a). The variation in the intensity of the reflected laser light as a function of membrane deflection is described by the following expression:

$$\Delta I_{laser} = \alpha(V_G) \sin\left(4\pi \cdot \frac{d(V_G)}{\lambda} + \varphi\right), \tag{S52}$$

where $d(V_G)$ represents the deflection of the suspended flake due to the applied gate voltage $V_G$, $\varphi$ is the initial phase, and $\lambda$ is the laser wavelength. The pre-factor $\alpha(V_G)$ accounts for the modifications in the optical properties of the 2D membrane under strain. Since the laser energy used in our experiments is approximately 0.6 eV away from the excitonic resonance of MoSe$_2$, strain-induced modulation of the band structure may cause minor changes in the optical constants. Thus, $\alpha(V_G)$ is approximated via expansion as $\alpha(V_G) = \alpha_0 + \alpha_1|V_G|$.

To relate the applied gate voltage $V_G$ to the membrane deflection $d$, we consider the system's total energy $U$, which is the sum of the electrostatic and elastic energy components:

$$U = U_{\text{elec}} + U_{\text{elas}}.$$

Minimizing the total energy $\frac{\partial U}{\partial d} = 0$, following the approach in Ref. [13], yields the implicit relation between membrane deflection and gate voltage:

$$\frac{256 E_Y t \cdot d^3(V_G)}{9L^2} + \frac{16 E_Y t \varepsilon_0 \cdot d(V_G)}{3} - \frac{1}{2} C_G V_G^2 = 0. \tag{S53}$$

Here $C_G$ is the capacitance of the system, $E_Y$ is the Young's modulus of the membrane, $t$ is its thickness, and $\varepsilon_0$ is the built-in (pre-)strain. The later can be found by comparing exciton emission energy in the regions of larger strain (center of the membrane) and regions with lower strain (supported on gold or edges of the membrane Fig. S1). By solving Eq. (S53), we obtain a fit function for the membrane deflection $d(V_G)$ as:

$$d(V_G) = \frac{\sqrt{h(V_G, a_1, a_2, \varepsilon_0)}}{3 a_2 \cdot 2^{\frac{1}{3}}} - \frac{a_1 \cdot 2^{\frac{1}{3}}}{\sqrt{h(V_G, a_1, a_2, \varepsilon_0)}}, \tag{S54}$$

where $a_1$ and $a_2$ are material specific parameters, and the function $h(V_G, a_1, a_2, \varepsilon_0)$ is defined as:

$$h(V_G, a_1, a_2, \varepsilon_0) = \left(\sqrt{|108 a_1^3 a_2^3 + (27 a_2^2 |V_G|^2 - 27 a_2^2 \varepsilon_0)^2|} - 27 a_2^2 \varepsilon_0 + 27 a_2^2 |V_G|^2\right)^{\frac{1}{3}}.$$

Using Eq. (S52) and Eq.(S54), we fit the laser interference data (Fig. S7b) to extract the membrane deflection, shown by the blue line in Fig. S7c. We then relate membrane deflection to the biaxial strain in the suspended monolayer using a parabolic membrane approximation [13]:

$$\varepsilon = \frac{1+\varepsilon_0}{2}\sqrt{1 + \frac{16d^2}{L^2}} + \frac{(1+\varepsilon_0)L}{8d}\log\left(\sqrt{1 + \frac{16d^2}{L^2}} + \frac{4d}{L}\right) - 1, \quad (S55)$$

where $L$ is the membrane diameter. The dependence of strain on gate voltage is presented in Fig. S7c, calculated using $L = 5.2\,\mu$m and $\varepsilon_0 = 0.2$. For small deflections, where strain is below 0.6%, the strain estimated from Eq. (S55) closely matches a simpler approximation $\varepsilon = 0.8\left(\frac{2d}{L}\right)^2$, with a discrepancy of less than 5% [14].

Next, using the extracted strain values, we determine the shift in the neutral exciton emission energy as a function of strain in MoSe$_2$ (Fig. S7d). The observed linear dependence of the emission energy on strain across the entire strain range confirms the accuracy of our strain measurements, as the conversion from the applied voltage to the induced strain is independent of the photoluminescence data. By fitting the data to a linear function, we find the parameter $A$ of the Eq. (1) in the main text equivalent to the slope $A = \Delta E/\Delta \varepsilon = 36 \pm 4\,\text{meV}/\%$, which agrees with previous studies [15, 16].

Finally, we relate the extracted biaxial strain to the uniaxial strain present in the sample. For that we define the uniaxiality of the strain as $U = \frac{\varepsilon_{xx}-\varepsilon_{yy}}{\varepsilon_{xx}+\varepsilon_{yy}}$. This expression connects the biaxial and uniaxial strain as $\varepsilon_{xx} - \varepsilon_{yy} = U \cdot 2(\varepsilon_{xx} + \varepsilon_{yy})$. To determine the uniaxiality, we performed finite element analysis of the strain distribution in the sample (Fig. S1). The results show that the uniaxiality of the strain is around 80% and weakly depends on the biaxial strain remaining nearly constant across the applied strain range. This modeling results can be approximated by simple analytical equation. We note that the strain along the major ($a$) and minor ($b$) axes of an ellipse is equivalent to the strain in circular membranes of the respective diameters. In a simple approximation, the strain in a circle is proportional to $\varepsilon \sim 1/D^2$, where $D$ is the diameter. Therefore, the ratio of strain along the major and minor axis is $\varepsilon_{xx}/\varepsilon_{yy} = (a/b)^2$. For a typical device with $a = 8\,\mu$m and $b = 3\,\mu$m, the uniaxiality is close to the results of finite element simulation: $U = \frac{(a/b)^2-1}{(a/b)^2+1} \approx 80\%$. Remarkably, the uniaxiality depends weakly on the exact ratio of the major and minor axes; increasing $U$ from 75% to 85% would require an approximately 25% increase in $a$ to 10.5 $\mu$m with $b$ fixed at 3 $\mu$m. We note that the exact shape of the membrane's edge as well as nonuniform distribution of the pre-strain may affect the uniaxiality of the strain.

# S3 Simulation of pseudo-Zeeman splitting at large pseudomagnetic fields

In this section, we provide details on the simulation of pseudo-Zeeman splitting used in the main text and compare the resulting behavior with experimental data. We show that thermal depletion of the higher-energy pseudo-Zeeman split state sufficiently explains the saturating dependence of the exciton splitting at high pseudomagnetic fields (Fig. 2d of the manuscript). We briefly summarize the experimental procedure and then describe the simulation approach that we used to best match the experiment.

In short our experiential procedure is the following. We use linearly polarized light to excite a pseudospin in a specific state on the Bloch sphere (see Fig. 1c of the main text). This pseudospin undergoes damped Larmor precession toward the pseudomagnetic field direction and eventually radiatively decays, imprinting its state onto the photoluminescence spectra. We then employ a linear polarizer (analyzer) to isolate the emission of a specific pseudospin state on the Bloch sphere. The pseudo-Zeeman splitting manifests as a difference in the emission energy of the two pseudospin states aligned along and opposite to the pseudomagnetic field.

To map this splitting, we measure the emission spectra of the sample as a function of the analyzer angle, keeping the excitation and detection polarizations parallel to reduce the influence of the pseudospin generation rate ($G$ in Eq. (S9)). Each spectrum is fitted with Gaussian functions to extract peak positions at each angle, and the resulting dependence is analyzed using $E = E_0 + \hbar\Omega/2 \cos\varphi$ to extract the pseudo-Zeeman splitting.

In our simulations, we aim to closely replicate the experimental procedures. First, we calculate the amount of pseudospin generated in the MoSe$_2$ sample using: $N = \frac{P\sigma\lambda 2\pi\tau}{\hbar c \pi r^2}$, where $P = 1\,\mu$W is the laser power, $\sigma = 5\%$ is the absorption coefficient, $\lambda = 680$ nm is the excitation wavelength, $r = 0.5\,\mu$m is the laser spot radius, and $\tau = 2$ ps is the exciton lifetime. The generation rate of pseudospin lies in the $xy$-plane, forming an angle $2\varphi$ with the pseudomagnetic field direction, and is set to 3% (see Fig. S3) of the total exciton density, $G = 0.03 N/\tau$.

We then calculate the steady-state pseudospin components using Eqs. (S9) with experiment informed parameter: $\tau = 1.8$ ps, $T_\parallel = 10$ ps, $T_{coh} = 0.8$ ps, and $T = 10$ K. Here we used the experimentally obtained parameters for the model with relaxation times independent of pseudomagnetic field [Eq. (S56), (S60)] to simplify the model. Subsequently, we compute the spectra at each analyzer angle $\varphi$ using Eq. (S51), accounting for an approximate 7% loss of polarization observed in these experimental settings. To model the unpolarized background likely arising from a tail of charged excitons (see spectra below 1.6 eV in Fig. S9a), we add a broad unpolarized Gaussian peak 22 meV below the exciton, with strain-dependent intensity reproducing the experimental behavior. The resulting data are then fitted with Gaussian functions to extract the peak positions, which are analyzed as a function of angle using the same equation as in the experiment.

The simulation results at a large pseudomagnetic field closely match the experimentally observed exciton splitting (Fig. S9a). Notably, while we used a pseudo-Zeeman splitting of 2.5 meV in the simulations, the exciton splitting extracted with method described above is approximately 0.8 meV. This discrepancy primarily arises from an underestimation of the emission energy of the higher-energy state due to thermal depletion (see Eq. (S28)). Although the extraction of splitting could potentially be improved by fitting two peaks for the split excitons, this approach would significantly reduce the stability of the procedure, given that the splitting is much less than the

width of the peaks.

We further compare the experimentally extracted splitting with the values obtained from simulations at different pseudomagnetic fields (points and solid line in Fig. S9b). The close agreement between the experiment and simulation supports the assumption of thermal depletion of the higher-energy state at high pseudomagnetic fields. Additionally, we compare the experimentally measured pseudospin polarization $S_\parallel$ with the simulation output (Fig. S9c). To satisfy experimental conditions, we conducted a separate set of simulations with a fixed direction of the generation term $G$. The simulated splitting increases linearly at low fields and saturates at high pseudomagnetic fields, consistent with the experimental data.

We note that while the model accurately describes the behavior of the pseudo-Zeeman splitting, it is qualitative in nature, as it involves simplified background emission.

## S4 Analysis of Pseudospin Time Constants

In this note, we apply the pseudospin model developed in Note S1 to analyze the pseudospin distribution under uniaxial strain. Our objective is to relate the experimentally measured dependence of pseudospin on the pseudomagnetic field to the pseudospin relaxation time constants $T_\parallel$ and $T_{coh}$.

**Longitudinal pseudospin component**

We begin by relating the model of pseudospin component parallel to the pseudomagnetic field $S_\parallel$ Eq. (S28) to the experimental observations. The $S_\parallel$ is measured as the difference in emission intensities polarized parallel and perpendicular to the strain axis, defined as:

$$S_\parallel = \frac{I_\parallel - I_\perp}{I_\parallel + I_\perp}$$

here, $I_\parallel$ and $I_\perp$ are the intensities of neutral exciton emission with the analyzer set parallel and perpendicular to the strain axis $x$, respectively.

On a first glance the first term in Eq. (S28) is independent of the pseudomagnetic field, which implies a constant offset in $S_\parallel$ as a function of field. However, the density of photogenerated excitons in a specific polarization state — described by $G_\parallel$ — depends on the excitation wavelength. This is because the strain effectively increases the energy detuning between the excitation wavelength and the exciton energy, which reduces pseudospin polarization of excitons [17, 18, 19]. Therefore, we use high energy of laser (2.32 eV) to minimize the effect of $G$ in the measurements of the $S_\parallel$. At low pseudomagnetic fields, $S_\parallel$ is less than 4%, primarily due to pre-strain effects (Fig. 3d of the main text). At higher fields, polarization retention becomes negligible, as confirmed by the independence of emission intensity on excitation polarization (Fig. S2c, Fig. S3e). Therefore, we further simplify Eq. (S28) by assuming $G_\parallel = 0$ and taking the absolute value of the pseudospin:

$$S_\parallel = \frac{\tau}{\tau + T_\parallel} \tanh\left(\frac{\hbar\Omega}{2k_B T}\right). \tag{S56}$$

Next, we consider the dependence of $T_\parallel$ on the pseudomagnetic field. According to Eq. (S8), the relaxation time along the field direction, $T_\parallel = 1/\Gamma_{xx}$, increases approximately quadratically with the field strength. Assuming that this time at zero pseudomagnetic field is equal to $\tau_{coh} = 1/((\Omega_k^{LT})^2 \tau_{sc})$, measured in four-wave mixing experiments [20, 21], we derive the final equation:

$$S_\parallel = \frac{1}{1 + \frac{\tau_{coh}}{\tau}\left(1 + \frac{\Omega^2}{\tau_{coh}^2(\Omega_k^{LT})^4} + \sqrt{\left(1 + \frac{\Omega^2}{\tau_{coh}^2(\Omega_k^{LT})^4} + \frac{1}{\tau_{coh}^4(\Omega_k^{LT})^4}\right)^2 - \frac{4\Omega^2}{\tau_{coh}^4(\Omega_k^{LT})^6}}\right)} \tanh\left(\frac{\hbar\Omega}{2k_B T}\right). \tag{S57}$$

We fitted the experimental data by treating the temperature $T$ and the root-mean-square exciton longitudinal–transverse splitting $\Omega^{LT} = \sqrt{\langle \Omega_k^{LT^2} \rangle}$ (where angular brackets denote the averaging over the Boltzmann distribution of excitons) as fitting parameters, while adopting the exciton lifetime $\tau$ and coherence time $\tau_{coh}$ from the literature: for $WSe_2$, $\tau = 2$ ps and $\tau_{coh} = 0.44$ ps; for $MoSe_2$, $\tau = 2$ ps and $\tau_{coh} = 0.52$ ps [20, 21, 22, 23, 24, 25, 26]. The fit yielded values of $\Omega^{LT} = 10.4 \pm 1.3$ T for $WSe_2$ and $\Omega^{LT} = 12.0 \pm 1.1$ T for $MoSe_2$ in reasonable agreement with theoretical

predictions, see Note S5. The exciton temperatures were determined to be $T = 13.1 \pm 1.1$ K for WSe$_2$ and $T = 22.4 \pm 1.9$ K for MoSe$_2$, consistent with previous reports for optically initialized excitons [27, 28].

Using Eq. (S7) and the extracted $\Omega^{\text{LT}}$, we estimated the characteristic scattering time $\tau_{\text{sc}} = 1/(\tau_{\text{coh}}(\Omega_k^{\text{LT}})^2)$. We found $\tau_{\text{sc}} = 0.76 \pm 0.11$ ps for WSe$_2$ and $0.49 \pm 0.05$ ps for MoSe$_2$, values that are consistent with the theoretical estimates in Ref. [3], see also Note S5. Here we assumed that scattering time is constant for the entire range of pseudomagnetic field, unintended changes of Fermi energy may lead to slight variation of this time. A comparison between the simpler model with fixed $T_\parallel$ [Eq. (S56)] and the more complex model [Eq. (S57)] is shown in Fig. S8. These results constitute the first measurement of exciton longitudinal–transverse splitting in transition metal dichalcogenides. All the parameters are summarized in the table below.

|  | $\tau$ | $\tau_{\text{coh}}$ | $\Omega^{\text{LT}}$ | $\tau_{\text{sc}}$ | $T$ |
|---|---|---|---|---|---|
| MoSe$_2$ | 2 ps | 0.52 ps | $12.0 \pm 1.1$ T | $0.49 \pm 0.05$ ps | $22.4 \pm 1.9$ K |
| WSe$_2$ | 2 ps | 0.44 ps | $10.4 \pm 1.3$ T | $0.76 \pm 0.11$ ps | $13.1 \pm 1.1$ K |

We note that the lifetime of neutral exciton population could be longer than the values adopted from the literature, as typical measurements like time-resolved photoluminescence are not sensitive to hot excitons outside light cone [29]. Additionally, doping can modify the exciton lifetime by introducing new recombination pathways [24], and cavity effects—such as interference between the suspended monolayer and the substrate—can influence the radiative decay rate [30]. However, our extracted exciton longitudinal–transverse splitting $\Omega^{\text{LT}}$ is fairly insensitive to variations in the exciton lifetime. For instance, when we double the exciton lifetime $\tau$ of MoSe$_2$ to 4 ps, we find that $\Omega^{\text{LT}}$ decreases by approximately 10% to 10.9 T, which remains within the experimental error of our measurement.

**Orthogonal pseudospin component**

Next, we analyze the pseudospin component along the $z$-axis under pseudomagnetic field in the $x$ direction when pseudospin along $y$-axis is initialized using linearly polarized light diagonal to the strain tensor axes. This component, $S_z$, describes the imbalance between $\sigma^+$ and $\sigma^-$ emissions, and it is measured as

$$S_z = \frac{I_{\sigma^+} - I_{\sigma^-}}{I_{\sigma^+} + I_{\sigma^-}}.$$

We find that at low pseudomagnetic field $S_z$ increases rapidly, and after $\sim 4$ T it starts to slowly decrease. In contrast other excitons show lower $S_z$ (Fig. S6). To compare this behaviour to theory use Eq. (S9b) with the result:

$$S_z = \frac{\tau_\perp^2 G \Omega}{1 + (\tau_\perp \Omega)^2}, \tag{S58}$$

where $\tau_\perp = 1/\sqrt{\tilde{\Gamma}_{yy} \tilde{\Gamma}_{zz}}$. This equation suggest that high initial polarization $G$ of excitons is required to observe significant pseudospin rotation. Therefore, neutral exciton in WSe$_2$ is an optimal system for observation of pseudospin effect due to large polarization retention even for excitation above the exciton resonance, unlike other excitonic states in this material or Mo-based materials [17, 18].

17

To quantitatively describe the dependence of $S_z$ on the pseudomagnetic field, we first need to factor out the dependence of $G$ on strain. To this end, we measure polarization-resolved photoluminescence (PL) of a circular sample, since in elliptical samples the pseudomagnetic field influences pseudospin dynamics (Fig. S3). Setting $\Omega = 0$ in Eqs. (S9), we find that the component of the pseudospin parallel to the field is $S^c = 2G\tau_\perp = G\tau_\|$, where $\tau_\| = 1/\tilde{\Gamma}_{xx}$, utilizing the relation $\Gamma_{zz} = 2\Gamma_{xx}$ at zero field [Eq. (S6)] and assuming that $\tau \gg \tau_\|, \tau_\perp$. We measure the linear polarization because the valley lifetime—which defines the retention of circular polarization (pseudospin along $z$) in WSe$_2$—is affected by dark excitons and resident carriers that can preserve the valley state for over 5 ps [31]; however, the coherence is lost on much shorter timescales ($< 0.5$ ps) [17, 18]. To measure $S^c$, we excited the sample with linearly polarized light and measured the intensity of emission parallel and perpendicular to the excitation polarization. Hence,

$$G\tau_\perp = \frac{S^c}{2} = \frac{I_{\|,exc} - I_{\perp,exc}}{2(I_{\|,exc} + I_{\perp,exc})}, \quad (S59)$$

where $I_{\|,exc}$ and $I_{\perp,exc}$ are the intensity of the neutral exciton emission parallel and perpendicular to the polarization of the excitation. From the experimental data, we find that the polarization retention $G\tau_\perp$ decays exponentially with the strain (Fig. S5), which we attribute to the increasing separation between the laser wavelength and the exciton energy.

Finally, we find that ratio of $S_z$ and $S^c$ is described by the analog of Eq. (S58) free of generation term $G$:

$$S_z^* = \frac{\tau_\perp \Omega}{1 + (\tau_\perp \Omega)^2}. \quad (S60)$$

The equation suggests that the orthogonal pseudospin component $S_z^*$ increases linearly at low pseudomagnetic fields, peaking at $\Omega \tau_\perp = 2\pi$, which corresponds to a full revolution of the pseudospin around the field during the transverse relaxation time. At higher fields, the pseudospin undergoes multiple rotations around the field $\Omega$, resulting in a decreased of assymetry between $\sigma^+$ and $\sigma^-$ emission. This behavior is consistent with the experimental data shown in Fig. 3 of the main text. Specifically, we obtained the data by dividing the pseudospin in the elliptical sample $S_z$ (Fig. S5a) by the value measured in the circular device $S^c$ (blue line in Fig. S5d). We then fitted the resulting $S_z^*$ to Eq. (S60).

It is noteworthy that the dependence of the transverse relaxation time on the pseudomagnetic field is less pronounced than that of the longitudinal relaxation time $T_\|$. According to Eq. S6, the relaxation rate along the $z$-axis is given by $\Gamma_{zz} = \Gamma_s \left(1 + \frac{1}{1+(\Omega \tau_{sc})^2}\right)$, where $\tau_{sc} = 1/(\tau_{coh}(\Omega_k^{LT})^2)$. Consequently, $T_{coh}$ increases from $T_{coh} = \tau_{coh}$ at low fields to $T_{coh} = 2\tau_{coh}$ at high fields. Fitting the data with a field independent $T_{coh}$ yields $T_{coh} = 0.8 \pm 0.2$ ps. In contrast, a more complex fit incorporates the effect of the depolarization field caused by the longitudinal-transverse splitting $\Omega_k^{LT}$, Eq. (S8):

$$S_z^* = \frac{\tau^* \Omega}{1 + (\tau^* \Omega)^2}, \quad \tau^* = \frac{\tau \tau_{coh} f}{\tau + \tau_{coh} f}, \quad f = 1 + \frac{\Omega^2/\tau_{coh}^2 (\Omega_k^{LT})^4}{1 + \sqrt{\left(1 + \frac{\Omega^2}{\tau_{coh}^2 (\Omega_k^{LT})^4} + \frac{1}{\tau_{coh}^4 (\Omega_k^{LT})^4}\right)^2 - \frac{4\Omega^2}{\tau_{coh}^4 (\Omega_k^{LT})^6}}} \quad (S61)$$

Here, the value of $\Omega_k^{LT}$ is determined by the dependence of $T_{coh}$ at high field, where the uncertainty is greater. Therefore, for the fitting procedure, we fixed $\Omega_k^{LT}$ to the value extracted from the previous subsection, 10.4 T.

Figure S8 shows the comparison of the two models. The main difference between the two fits manifests as a slight divergence at high field strengths.

## S5 Estimate of the exciton LT splitting

For a free monolayer the following expression for the neutral exciton LT splitting holds [32, 33, 34]

$$\Omega_k^{LT} = \frac{\Gamma_0}{q} k, \tag{S62}$$

where $\Gamma_0$ is the exciton radiative decay rate (at the in-plane wavevector $k = 0$, i.e., in the radiative cone), $q = \omega_0/c$ is the wavevector of photon emitted by the exciton, $\omega_0 = (E_g - E_B)/\hbar$ is the exciton transition frequency with $E_g$ and $E_B$ being the free-particle bandgap and exciton binding energy, respectively. Equation (S62) holds for $k \gg q$. In the presence of the screening induced, e.g., by the encapsulation and substrate one has instead of Eq. (S62)

$$\Omega_k^{LT} = \frac{\Gamma_0}{\varepsilon q} k, \tag{S63}$$

where $\varepsilon$ is the effective dielectric constant of the surrounding, see Ref. [35] for details.

We are interested in $\Omega_k^{LT}$ averaged over the thermal distribution of excitons (this quantity enters the spin decay rates for thermalized excitons):

$$\bar{\Omega}_k^{LT} = \left[ \int_0^\infty \frac{dE_k}{k_B T} e^{-E_k/(k_B T)} \left(\Omega_k^{LT}\right)^2 \right]^{1/2} = \frac{\Gamma_0}{\varepsilon q} \sqrt{\frac{2 M k_B T}{\hbar^2}}, \tag{S64}$$

where $M$ is the exciton translational motion mass and $T$ is the temperature.

Taking $\tau_r = 1/(2\Gamma_0) = 1$ ps, $M = 0.75 m_0$ ($m_0$ is free-electron mass), $T = 20$ K, $\varepsilon = 1$, and $E_g - E_B = 2$ eV we have

$$\bar{\Omega}_k^{LT} \approx 9 \text{ ps}^{-1}, \quad \hbar \bar{\Omega}_k^{LT} \approx 6 \text{ meV}, \quad \frac{\hbar \bar{\Omega}_k^{LT}}{2\mu_B} \approx 52 \text{ T}.$$

This sizable field corresponds to completely clean monolayer. The discrepancy between the estimate and the experimentally measured value may arise due to underestimation of dielectric constant or radiative lifetime [30]. For instance, residual layers of polydimethylsiloxane (PDMS) related to our fabrication technique, can increase $\varepsilon$ to around 2.5 [36], reducing the effective field to $\Omega_k^{LT} \approx 21$ T.

At acoustic phonon scattering

$$\frac{1}{\tau_{sc}^{ac}} = C \frac{k_B T}{\hbar}, \tag{S65}$$

with the material-dependent coefficient $C = 0.5 \ldots 1$ [37, 38, 39] we have at $T = 20$ K (and $C = 1$)

$$\bar{\Omega}_k^{LT} \tau_{sc}^{ac} \approx 3.5 \text{ (@ } \varepsilon = 1\text{)} \quad \text{or} \quad \approx 1.3 \text{ (@ } \varepsilon = 2.5\text{)}.$$

Hence, the product $\bar{\Omega}_k^{LT} \tau_{sc}^{ac}$ is not small, typically. Note that $\hbar \bar{\Omega}_k^{LT}$ is comparable with the mean kinetic energy of excitons which generally complicates the spin dynamics [40, 41].

## S6 Calculation of Fermi energy

In this supplementary note, we determine the Fermi energy in suspended devices using two complementary methods. The first method is based on the dependence of the Fermi polaron-polariton binding energy on the applied gate voltage, while the second method uses the standard capacitor model to cross-validate the results.

The relationship between the binding energy of a Fermi polaron and the carrier density is well-established in the literature [42, 43, 44]. Briefly, the Fermi polaron is a quasiparticle formed by an exciton interacting with a Fermi sea of carriers, where the density of carriers governs the binding energy. This dependence is described by the equation:

$$E_{FP} = E_{FP}^0 + kE_F,$$

where $E_{FP}^0$ is the binding energy of the Fermi polaron at zero carrier density, and $k$ is a constant. In our study, we use values that were measured for hBN encapsulated monolayers, $k = 1.1$, with $E_{FP}^0 = 26.1$ meV [43, 44]. The Fermi energy $E_F$, extracted as a function of applied gate voltage $V_G$, exhibits a near-linear dependence, as shown in Fig. S3f, consistent with theoretical expectations. Note that the binding energy of the FP in suspended samples $E_{FP}^0$ may differ from reported values due to differences in the dielectric environment for the suspended sample.

The carrier density $n_e$ is related to the Fermi energy by the equation:

$$n_e = \frac{E_F m_e}{\pi \hbar^2},$$

where we assumed that the two valleys are occupied, and $m_e \approx 0.5 m_0$ is the electron effective mass, with $m_0$ being the free electron rest mass. For practical calculations, the relationship simplifies to:

$$n_e \approx 0.2 \times 10^{12} E_F, \tag{S66}$$

where units for $n_e$ is cm$^{-2}$ and $E_F$ is meV.

This method assumes that the exciton-FP separation depends solely on doping; however, strain induced by the applied gate voltage may also affect the separation. Thus, a second method is required to validate the results.

We verify the results using a capacitor model to estimate the carrier density induced by the gate voltage. The carrier density $n_{e,h}$ in the membrane is approximated using a parallel plate capacitor model:

$$n_{e,h} = \frac{V_G \epsilon_0}{e} \left( \frac{\epsilon_{SiO_2}}{d_{SiO_2} + \epsilon_{SiO_2}[d_{Au} - d(V_G)]} \right),$$

where $\epsilon_0$ is the vacuum permittivity, $\epsilon_{SiO_2} = 3.6$ is the dielectric constant of SiO$_2$, $e$ is elementary charge, $d_{SiO_2} = 1000$ nm is the SiO$_2$ thickness, $d_{Au} = 500$ nm is the distance between the gold surface and SiO$_2$, and $d(V_G)$ is the deflection of the membrane, determined via interferometry, see Note. S2. This model accounts for the increasing capacitance due to the deflection of the suspended flake. The induced carrier density can then be converted to Fermi energy using Eq. (S66).

Both methods show strong agreement (Fig. S3f), validating the accuracy of the exciton-FP splitting approach for determining the Fermi energy.

# S7 Evaluation of Fermi polaron pseudo-Zeeman splitting

The theoretical curves presented in Fig. 4d of the main text are derived following Eqs. (31) and (33) of Ref. [45]. For MoSe$_2$, the ratio between the neutral exciton and Fermi polaron pseudo-Zeeman splittings is given by

$$\frac{\hbar\Omega_{FP}}{\hbar\Omega_X} = \frac{(M_X/M_T)^3}{4\sinh^2[\frac{1}{2}(M_X/M_T)^2]}\frac{E_F}{E_b}, \tag{S67}$$

where $M_X$ and $M_T = \frac{3}{2}M_X$ denote the exciton and Fermi polaron masses, respectively, and $E_b = 26.1\,\text{meV}$ is the binding energy relative to the exciton. Note that the equation is valid only at $E_F \ll E_b$.

In WSe$_2$, the Fermi polaron splitting is described by

$$\frac{\hbar\Omega_{FP}}{\hbar\Omega_X} = \frac{(M_X/M_T)^3}{2\sinh^2[\frac{1}{2}(M_X/M_T)^2]}\frac{E_F}{\Delta}, \tag{S68}$$

where $\Delta = 6.8\,\text{meV}$ is the energy difference between intra- and inter-valley trion states [44]. Since $\Delta \ll E_b$ the FP splitting in WSe$_2$ is larger for the same field strength and Fermi energy.

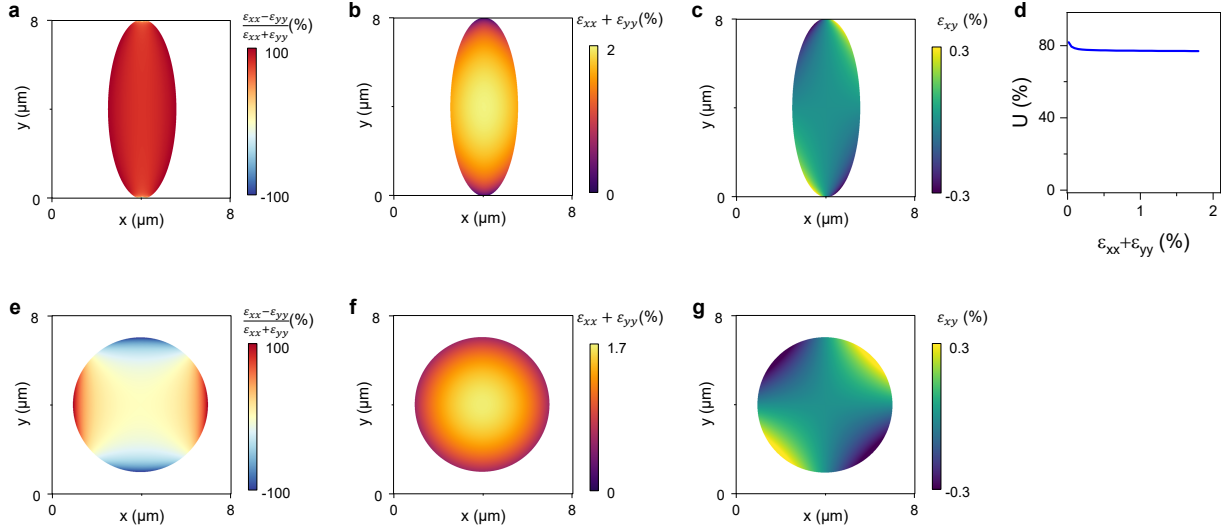

Figure S1: **Strain profile in elliptical and circular devices. a-c**: False color maps of strain uniaxiality $\frac{\varepsilon_{xx}-\varepsilon_{yy}}{\varepsilon_{xx}+\varepsilon_{yy}}$, total strain ($\varepsilon_{xx} + \varepsilon_{yy}$), and shear strain ($\varepsilon_{xy}$) in an elliptical device simulated using finite element analysis. Simulations were performed using COMSOL Multiphysics software, employing the electrostatics and structural mechanics modules. Using the electrostatic module we calculate the electrostatic pressure on the membrane, which was then applied in the structural mechanics module to simulate the membrane's response. The parameters used in the simulation of the TMD membrane were set to a thickness of 0.7 nm, Poisson's ratio of 0.19, density of 9.32 g/cm$^3$, and Young's modulus of 117 GPa [46, 47]. Pretension was applied as an in-plane initial stress of 15 MPa, resulting in a strain of $\varepsilon_0 \sim 0.1\%$, consistent with experimental values. The total strain $\varepsilon_{xx} + \varepsilon_{yy}$ is homogeneous at the center of the device, with variation less than $\frac{\Delta\varepsilon}{\varepsilon} < 10\%$ within a 1 $\mu$m diameter region. In principle, shear strain may contribute to a more complex pseudomagnetic field (see Note S1), simulations indicate it is negligible at the device center. Note that the color scale in graph (a) differs from that in Fig. 1f of the main text for better comparison with the circular device. **d**: Dependence of strain uniaxiality $U$ on total strain at the center of the elliptical sample. Throughout the range of applied strain, the uniaxiality remains approximately 80%. **e-g**: False color maps of strain uniaxiality, total strain, and shear strain in a circular device. As expected from the device's symmetry, the uniaxiality of strain is zero at the device center.

23

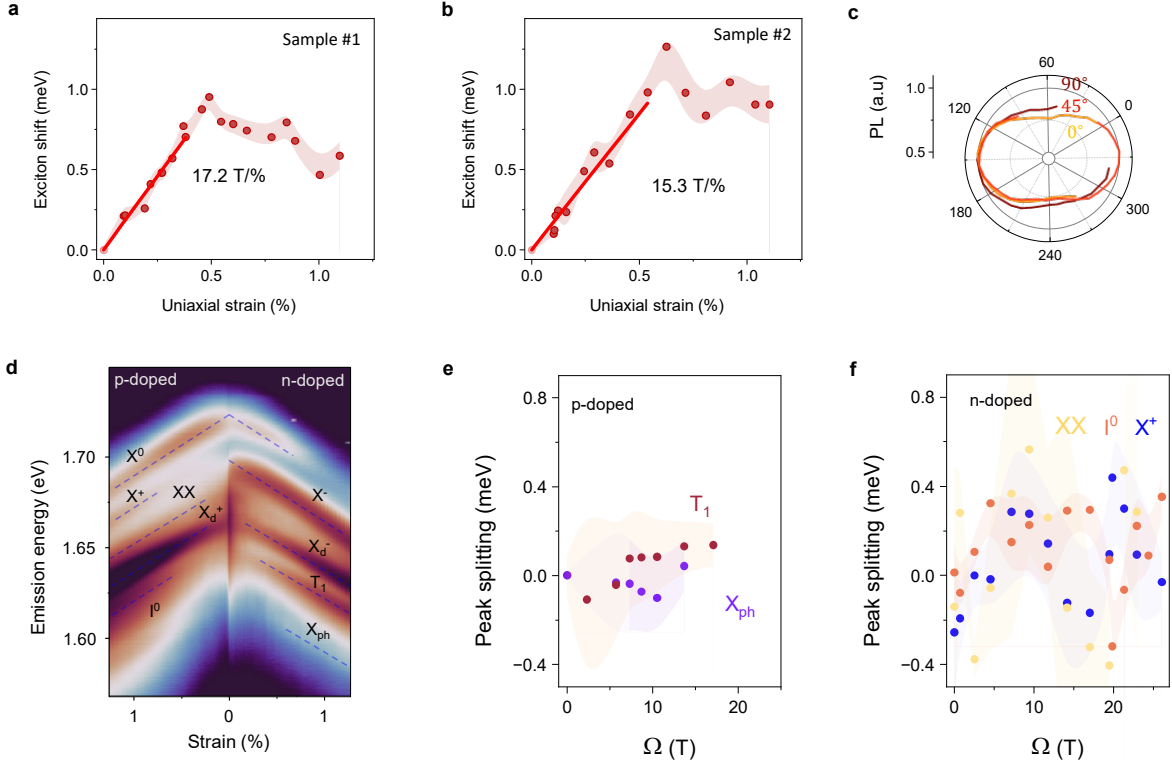

Figure S2: **Extended analysis of WSe$_2$ device. a,b**: Shift of the exciton emission energy in WSe$_2$ devices #1 (Fig. 3b,c of the main text) and #2 (Fig. 3d and Fig. 4 of the main text) as a function of applied uniaxial strain. The exciton shift is derived from the dependence of exciton emission energy on analyzer angle, following the procedure outlined in the main text. The strain-to-pseudomagnetic field conversion factor is extracted from a linear fit to the linear region of the data and determined to be $16.1 \pm 1.8$ T/%. Note that the largest uniaxial strain obtained in the WSe$_2$ devices is $\sim 1.5$%, but intensity of neutral exciton emission is too low to extract the exciton splitting due to effect of doping. **c**: Dependence of the neutral exciton emission intensity in WSe$_2$ on analyzer angle for a 2.32 eV excitation polarized at $\sim 0°$, $45°$, and $90°$ relative to the pseudomagnetic field (yellow, orange, and red lines, respectively). The pseudomagnetic field favors emission polarized along the $\sim 150°$ direction, coinciding with the minor axis of the ellipse (which is the direction of the $\Omega$). The emission pattern shows minimal dependence on excitation polarization, indicating low $G$, as the laser energy is significantly detuned from the exciton energy. **d**: False-color PL map of the strained elliptical sample for both polarities of the gate voltage ($V_G$), plotted on a logarithmic scale. In the negative voltage region (p-doped), we observe the neutral exciton (X$^0$), positively charged exciton (X$^+$), biexciton (XX), dark charged exciton (X$_d^+$), and the I$^0$ peak, commonly attributed to phonon replica of the intervalley dark exciton. For positive voltage (n-doped), additional features such as the negatively charged bright and dark excitons (X$^-$, X$_d^-$), the $T_1$ peak (origin debated), and the phonon replica peak (X$_{ph}$) are identified. Due to broad linewidths, only one X$^-$ peak is observed, likely a combination of singlet and triplet states, dominated by the brighter triplet. The strain-induced energy shift is symmetric with respect to $V_G$ polarity. **e-f**: Peak splitting of excitons omitted in the main text in p-doped and n-doped regimes, respectively. In the p-doped regime, negligible splitting is observed for the X$_{ph}$ peak, a finite splitting of the $T_1$ peak is detected, though it remains comparable to noise. In the n-doped regime, splitting of the XX, I$^0$, and X$^+$ peaks exhibit larger noise due to their low intensity relative to X$_d^+$, alongside significant peak broadening.

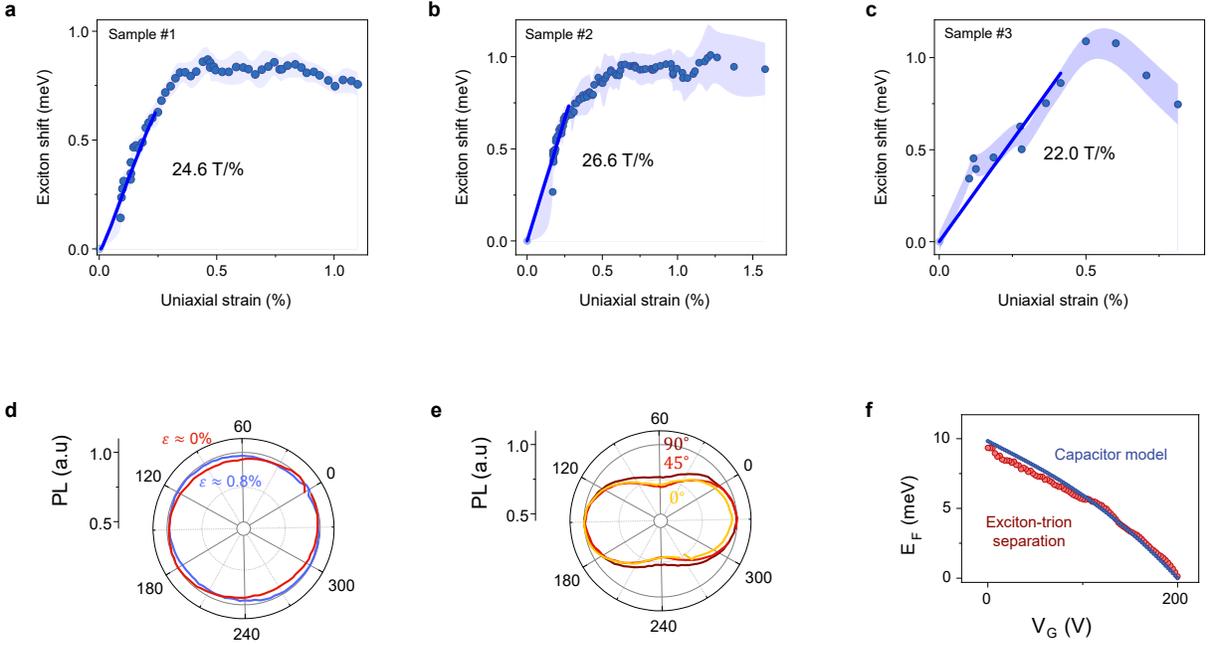

Figure S3: **Extended analysis of MoSe$_2$ device. a-c**: Shift of the exciton emission energy in MoSe$_2$ devices #1 (Fig. 2, 3, 4d of the main text), #2, and #3 as a function of applied uniaxial strain. The exciton shift is extracted from the dependence of exciton emission energy on analyzer angle, following the procedure outlined in the main text. The strain-to-pseudomagnetic field conversion factor is obtained from a linear fit to the linear region of the data. The average conversion factor is found to be $24.6 \pm 2.5$ T/%. **d**: Dependence of neutral exciton emission intensity in a circular MoSe$_2$ membrane on analyzer angle at 0% applied strain (red) and at $\varepsilon_{xx} = \varepsilon_{yy} = 0.8\%$. A slight dependence of emission intensity on the analyzer angle is observed at zero strain, likely due to the retention of initial polarization. However, the lack of angular dependence at finite biaxial strain in circular device confirms that pseudospin polarization in elliptical sample is induced by the pseudomagnetic field. **e**: Dependence of neutral exciton emission intensity in an elliptical MoSe$_2$ membrane on analyzer angle for 1.83 eV excitation polarization at $\sim 0°$, $45°$, and $90°$ relative to the pseudomagnetic field (yellow, orange, and red lines, respectively). The pseudomagnetic field favors emission polarized along the $\sim 150°$ direction, which coincides with the minor axis of the ellipse. The emission pattern shows minimal dependence on the excitation polarization, indicating low retention of initial polarization, as the laser energy is well detuned from the exciton energy. **f**: Comparison of Fermi energy dependence on gate voltage ($V_G$) extracted using exciton-polaron energy separation (red dots) and a capacitor model (blue points). For exciton-polaron separation, the model developed in ref.[42, 43, 44] is employed (see Note S6). To cross-check the results, an independent method—based on a capacitor model—is used (Note S6). Both methods yield consistent results, confirming the validity of the approaches.

25

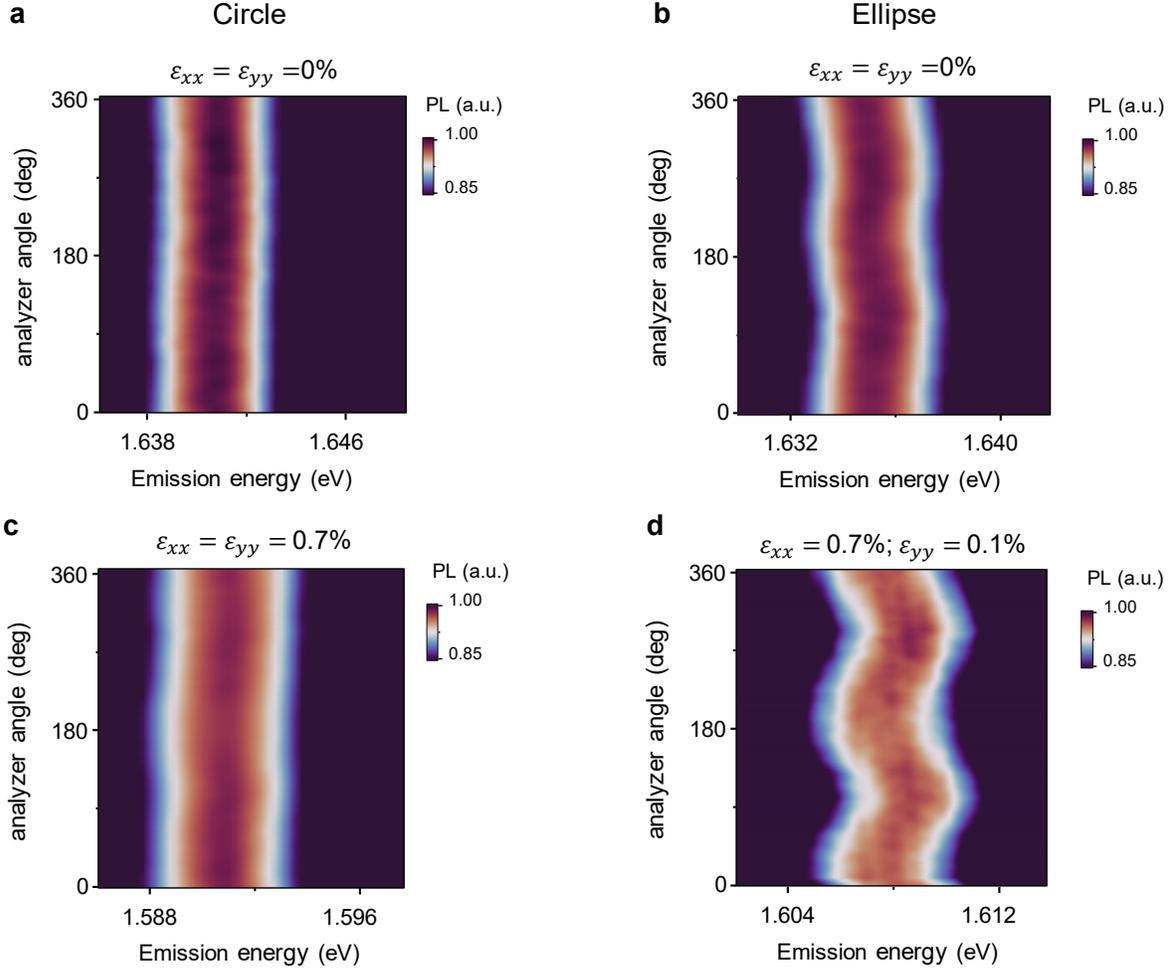

Figure S4: **Biaxial vs. uniaxial strained MoSe$_2$ a-b:** False-color map of neutral exciton PL in a circular and elliptical MoSe$_2$ membrane as a function of analyzer angle at 0% applied strain. Some dependence of emission intensity on the analyzer angle is assigned to prestrain that is estimated by comparison to the emission at the sample edge $\varepsilon_0 = 0.08\%$ in the elliptical sample. **c-d:** False-color map of neutral exciton PL in the same two devices under finite strain. This results in strain tensor $\varepsilon_{xx} = \varepsilon_{yy} = 0.7\%$ in circular membrane and $\varepsilon_{xx} = 0.7\%; \varepsilon_{yy} = 0.1\%$ in elliptical one. The emission intensity in the circular sample shows no dependence on the analyzer angle, while the elliptical sample exhibits a clear angular dependence, indicating the presence of a pseudomagnetic field. The pseudomagnetic field is oriented along the minor axis of the ellipse, as expected.

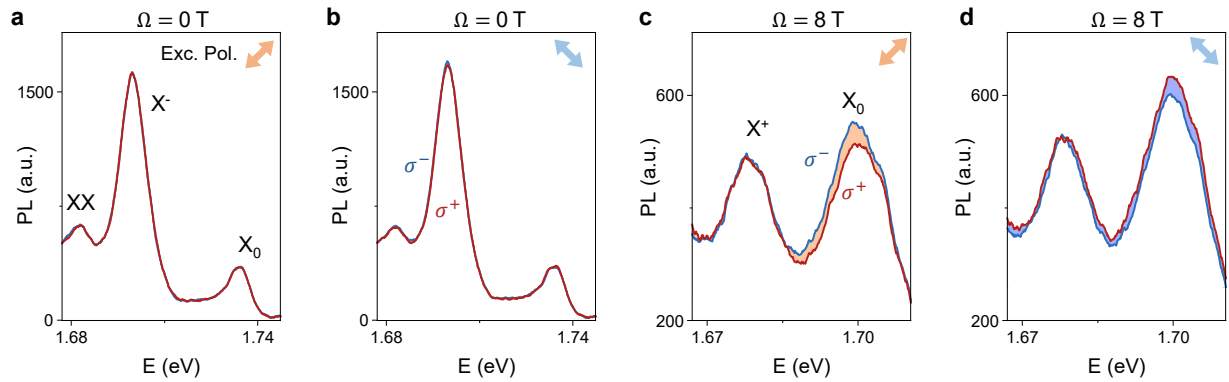

Figure S5: **Emergence of orthogonal pseudospin component in pseudomagnetic field a,b:**. Polarization resolved PL spectra of pristine WSe$_2$ monolayers under linearly polarized excitation at angles of +45° (a) and 45° (b) relative to the strain axis. The spectra exhibit negligible variation in emission intensity between $\sigma^+$ (red) and $\sigma^-$ (blue) polarized light, suggesting absence of pseudospin rotation in the absence of external pseudomagnetic fields. **c,d:** Polarization resolved PL spectra under an applied pseudomagnetic field of $\Omega = 8$ T, showing significant asymmetry in $\sigma^+$ and $\sigma^-$ polarized emission. Reversing the excitation polarization between +45° (c) and -45° (d) results in a sign flip of the asymmetry, consistent with pseudospin rotation induced by the pseudomagnetic field. Notably, slight differences in the spectra for the two excitation polarizations are observed; to mitigate setup-related effects, we average the extracted pseudospin values $S_z$ across both excitation polarizations.

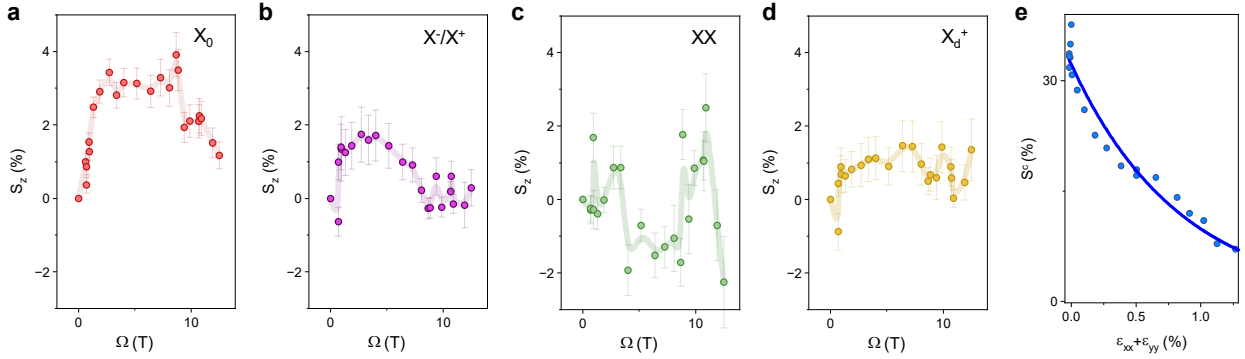

Figure S6: **Dependence of $S_z$ of $X_0$, $X^-$, XX, $X_d^+$ on pseudomagnetic field a-d:** Orthogonal pseudospin component $S_z$ of neutral excitons ($X^0$), bright charged excitons ($X^-/X^+$), biexciton (XX), and dark charged excitons ($X_d^+$) in WSe$_2$ as a function of the pseudomagnetic field. The solid splines serve as a guide to the eye. The largest signal is observed for neutral exciton, reflecting its higher generation $G$ (see Eq. (S58)). We also observe non-zero $S_z$ for charged excitons that peaks at lower pseudomagnetic field compared to neutral excitons. The study of it is complicated due to doping related change of $G$. Other species show a smaller signal close to the noise floor. We note that the device used in the measurement had slight initial doping, leading to a transition from negative trion to positive trion at around $\Omega \approx 2.5$ T. **e:** Pseudospin of neutral exciton $S^c = \frac{I_{\parallel,exc} - I_{\perp,exc}}{(I_{\parallel,exc} + I_{\perp,exc})}$ (Note S4) in a circular WSe$_2$ membrane (points). The reduction in pseudospin value is primarily due to the increasing separation between the laser wavelength and the exciton energy[19, 17], which we fit to an exponential decaying function (solid line). The measurement was conducted in a circular device to avoid effects of pseudomagnetic fields on pseudospin.

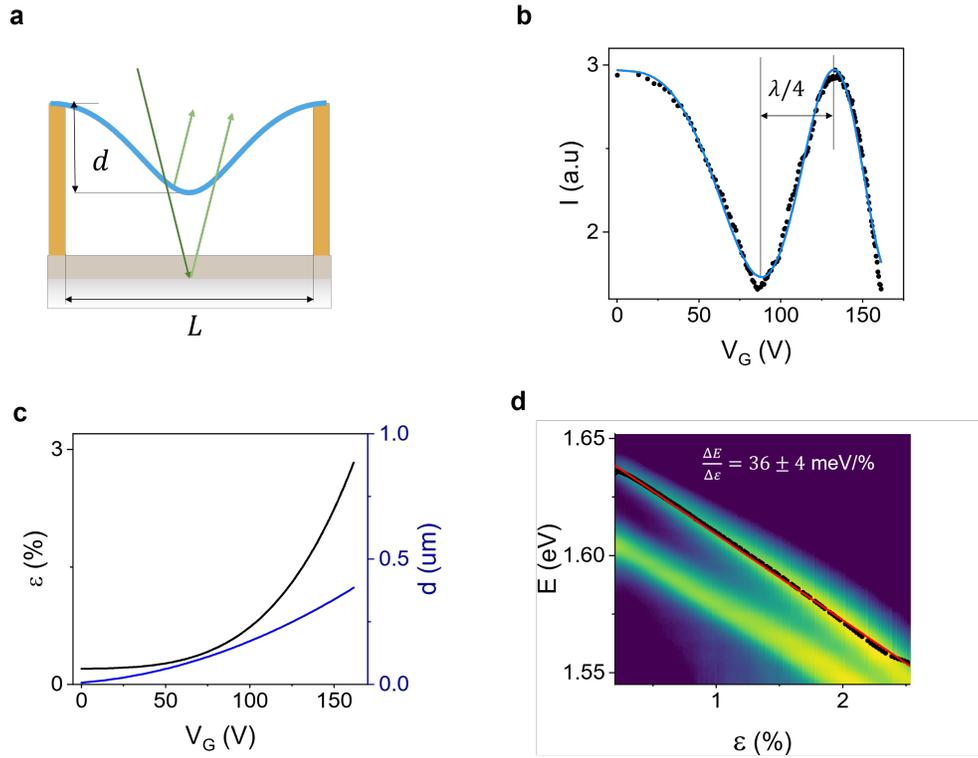

Figure S7: **Strain calibration a:** Schematic of the experimental setup used to measure strain in the device. The phase difference between the laser reflected off the membrane and the Si substrate is sensitive to membrane deflection, leading to optical interference. **b:** Deflection of the membrane as a function of gate voltage (dots) fitted to the theoretical model (blue line). The intensity modulation of reflected light corresponds to deflections equal to multiples of the laser wavelength. **c:** Deflection of the membrane (right axis, blue line) and strain $\varepsilon_{xx} + \varepsilon_{yy}$ (left axis, black line) as a function of gate voltage. **d:** False-color map of photoluminescence spectra of a MoSe$_2$ monolayer as a function of applied strain. The relationship between the excitonic energy shift and strain is determined by a linear fit (red line) to the shift of neutral exciton emission energy (black dots).

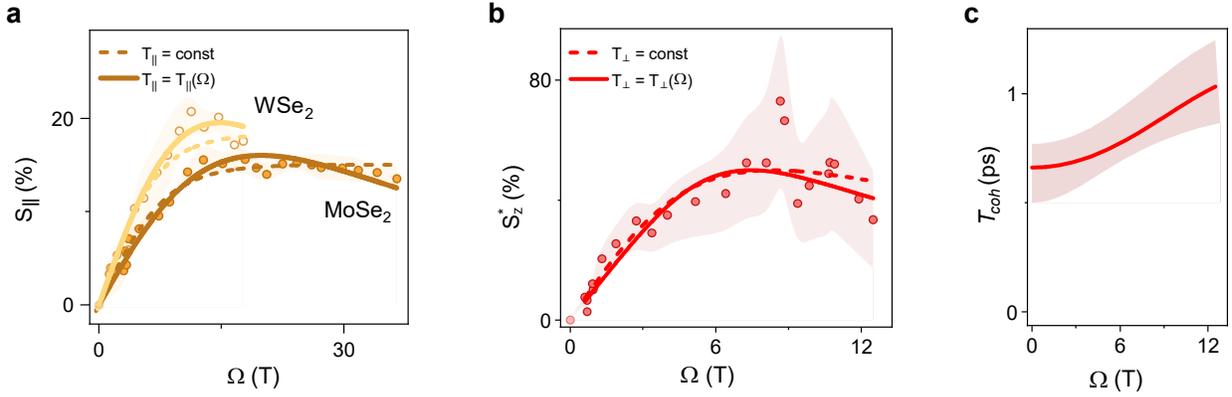

Figure S8: **The effect of pseudomagnetic field on relaxation times a:** The component $S_\parallel$ of the pseudospin as a function of the pseudomagnetic field in MoSe$_2$ and WSe$_2$ (filled and empty circles, respectively), along with fits to the simple model Eq. (S56) with fixed $T_\parallel$ (dashed lines) and the more complex model (S57) where $T_\parallel$ is a function of the field. The simple model yields $T_\parallel = 10.2$ ps for MoSe$_2$ and $T_\parallel = 8.9$ ps for WSe$_2$, using $\tau = 1.8$ ps and $\tau = 2$ ps, respectively, and $T \approx 5$ K. **b:** The component $S_z$ of the pseudospin as a function of the pseudomagnetic field strength in WSe$_2$, along with fits to the simple model Eq. (S60) with fixed $T_{coh}$ (dashed lines) and the more complex model Eq. (S61) where $T_{coh}$ is a function of the field. **c:** The dependence of the transverse relaxation time $T_{coh}$ on pseudomagnetic field strength. The slightly higher value of $T_{coh} = 0.68$ ps at low field compared to resultes of four wave mixing experiments at zero field $\tau_{coh} = 0.44$ ps [17, 18] could be related to preservation of valley degrees of freedom by dark excitons and resident carriers, effectively prolonging the lifetime of the pseudospin along the z-axis.

30

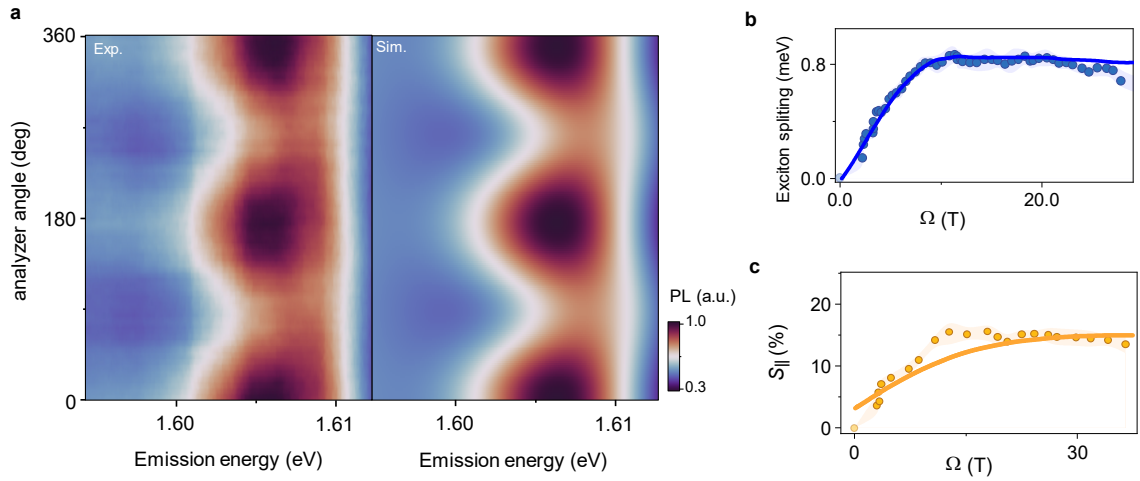

Figure S9: **Zeeman splitting at high pseudomagnetic field a** Experimentally measured and simulated false-color map of neutral exciton photoluminescence in MoSe$_2$ as a function of analyzer angle under a pseudomagnetic field of 21 T normalized to global maximum. Despite the underlying Zeeman splitting of approximately 2.5 meV, the splitting extracted using the procedure described in the main text yields approximately 0.8 meV in both experimental and simulated maps. **b** Extracted exciton splitting versus pseudomagnetic field for experimental data (points) and independent simulation (solid line). **c** Parallel component of pseudospin $S_\parallel$ versus pseudomagnetic field for experimental data (points) and independent simulation (solid line).

[29] Lagarde, D. *et al.* Efficient electron spin relaxation by chiral phonons in WSe 2 monolayers. *Physical Review B* **110**, 195403. URL https://link.aps.org/doi/10.1103/PhysRevB.110.195403.

[30] Fang, H. *et al.* Control of the exciton radiative lifetime in van der waals heterostructures. *Physical review letters* **123**, 067401 (2019).

[31] Raiber, S. *et al.* Ultrafast pseudospin quantum beats in multilayer WSe2 and MoSe2. *Nature Communications* **13**, 4997 (2022). URL https://www.nature.com/articles/s41467-022-32534-3. Number: 1 Publisher: Nature Publishing Group.

[32] Glazov, M. M. *et al.* Exciton fine structure and spin decoherence in monolayers of transition metal dichalcogenides. *Phys. Rev. B* **89**, 201302 (2014). URL http://dx.doi.org/10.1103/PhysRevB.89.201302.

[33] Yu, H., Liu, G.-B., Gong, P., Xu, X. & Yao, W. Dirac cones and Dirac saddle points of bright excitons in monolayer transition metal dichalcogenides. *Nat Commun* **5**, 3876 (2014). URL http://dx.doi.org/10.1038/ncomms4876.

[34] Iakovlev, Z. A. & Glazov, M. M. Longitudinal-transverse splitting and fine structure of Fermi polarons in two-dimensional semiconductors. *J. Lumin.* **273**, 120700 (2024). URL https://www.sciencedirect.com/science/article/pii/S0022231324002643.

[35] Prazdnichnykh, A. I. *et al.* Control of the exciton valley dynamics in atomically thin semiconductors by tailoring the environment. *Phys. Rev. B* **103**, 085302 (2021). URL https://link.aps.org/doi/10.1103/PhysRevB.103.085302.

[36] Mark, J. E. Polymer data handbook. *(No Title)* (2009).

[37] Selig, M. *et al.* Excitonic linewidth and coherence lifetime in monolayer transition metal dichalcogenides. *Nature Communications* **7**, 13279 (2016). URL http://dx.doi.org/10.1038/ncomms13279.

[38] Shree, S. *et al.* Observation of exciton-phonon coupling in MoSe$_2$ monolayers. *Phys. Rev. B* **98**, 035302 (2018). URL https://link.aps.org/doi/10.1103/PhysRevB.98.035302.

[39] Glazov, M. M. Quantum interference effect on exciton transport in monolayer semiconductors. *Phys. Rev. Lett.* **124**, 166802 (2020). URL https://link.aps.org/doi/10.1103/PhysRevLett.124.166802.

[40] Grimaldi, C. Electron spin dynamics in impure quantum wells for arbitrary spin-orbit coupling. *Phys. Rev. B* **72**, 75307 (2005).

[41] Averkiev, N. S. & Glazov, M. M. Specific features of optical orientation and relaxation of electron spins in quantum wells with a large spin splitting. *Semiconductors* **42**, 958–966 (2008). URL https://doi.org/10.1134/S1063782608080149.

[42] Glazov, M. M. Optical properties of charged excitons in two-dimensional semiconductors. *The Journal of Chemical Physics* **153**, 034703 (2020). URL http://arxiv.org/abs/2004.13484. 2004.13484[cond-mat].


\fi

\end{document}